\documentclass[a4paper,11pt]{article}
\pdfoutput=1 

\usepackage{jcappub}
\usepackage{graphicx}
\usepackage{dcolumn}
\usepackage{amssymb,amsmath,bm,bbold}
\usepackage{color}
\usepackage[dvipsnames,table]{xcolor}
\usepackage{xfrac}
\usepackage{aas_macros}
\usepackage{mathrsfs}
\usepackage{subcaption}
\usepackage{rotating}
\usepackage{chngcntr}
\usepackage{lipsum}
\usepackage{units}
\usepackage{multirow}
\usepackage{xspace}
\usepackage{xfrac}
\usepackage{mathtools}
\usepackage{comment}
\usepackage{float}

\newcommand{\redmagic}{\textsc{redMaGiC}}
\newcommand{\mcal}{\textsc{Metacalibration}}
\newcommand{\nmt}{\textsc{NaMaster}}
\newcommand{\abacus}{\textsc{AbacusSummit}}
\newcommand{\planck}{{\it Planck}}
\newcommand{\nv}{\hat{\bf n}}

\newcommand{\hMpc}{$h\,{\rm Mpc}^{-1}$}
\newcommand{\iMpc}{${\rm Mpc}^{-1}$}

\definecolor{internationalkleinblue}{rgb}{0.0, 0.18, 0.65}
\hypersetup{urlcolor=internationalkleinblue, linkcolor=internationalkleinblue, citecolor=internationalkleinblue}

\usepackage[T1]{fontenc} 
\usepackage{natbib}
\title{Hefty enhancement of cosmological constraints from the DES Y1 data using a Hybrid Effective Field Theory approach to galaxy bias}
\author[a,1]{Boryana Hadzhiyska,}
\author[b]{Carlos Garc\'ia-Garc\'ia,}
\author[b]{David Alonso,}
\author[c]{Andrina Nicola,}
\author[d]{An\v ze Slosar}
\affiliation[a]{Harvard-Smithsonian Center for Astrophysics, 60 Garden St., Cambridge, MA 02138, USA}
\affiliation[b]{Department of Physics, University of Oxford, Denys Wilkinson Building, Keble Road, Oxford OX1 3RH, United Kingdom}
\affiliation[c]{Department of Astrophysical Sciences, Princeton University, Peyton Hall, Princeton, NJ 08544, USA}
\affiliation[d]{Brookhaven National Laboratory, Physics Department, Upton, NY 11973, USA}
\emailAdd{boryana.hadzhiyska@cfa.harvard.edu}

\abstract{We present a re-analysis of cosmic shear and galaxy clustering from first-year Dark Energy Survey data (DES Y1), making use of a Hybrid Effective Field Theory (HEFT) approach to model the galaxy-matter relation on weakly non-linear scales, initially proposed in \cite{1910.07097}. This allows us to explore the enhancement in cosmological constraining power enabled by extending the galaxy clustering scale range typically used in projected large-scale structure analyses. Our analysis is based on a recomputed harmonic-space data vector and covariance matrix, carefully accounting for all sources of mode-coupling, non-Gaussianity and shot noise, which allows us to provide robust goodness-of-fit measures. We use the \textsc{AbacusSummit} suite of simulations to build an emulator for the HEFT model predictions. We find that this model can explain the galaxy clustering and shear data up to wavenumbers $k_{\rm max}\sim 0.6\, {\rm Mpc}^{-1}$. We constrain $(S_8,\Omega_m) = (0.786\pm 0.020,0.273^{+0.030}_{-0.036})$ 
at the fiducial $k_{\rm max}\sim 0.3\, {\rm Mpc}^{-1}$,
improving to $(S_8,\Omega_m) = (0.786^{+0.015}_{-0.018},0.266^{+0.024}_{-0.027})$ at $k_{\rm max}\sim 0.5\, {\rm Mpc}^{-1}$. 
This represents a $\sim10\%$ and $\sim35\%$ improvement on the constraints derived respectively on both parameters using a linear bias relation on a reduced scale range ($k_{\rm max}\lesssim0.15\,{\rm Mpc}^{-1}$), in spite of the 15 additional parameters involved in the HEFT model.
We investigate whether HEFT can be used to constrain the Hubble parameter and find $H_0= 70.7_{-3.5}^{+3.0}\,{\rm km}\,s^{-1}\,{\rm Mpc}^{-1}$. 
Our constraints are investigative and subject to certain caveats discussed in the text.
}
\begin{document}
\maketitle
\flushbottom

  \section{Introduction}\label{sec:intro}
    We are entering the prime time era of photometric cosmological surveys. A number of these surveys have reported results over the past few years, including the Dark Energy Survey (DES)\footnote{\url{https://www.darkenergysurvey.org}.} \citep{1708.01530}, the Hyper Suprime-Cam survey (HSC)\footnote{\url{https://hsc.mtk.nao.ac.jp/ssp}.} \citep{1809.09148} and the Kilo-Degree Survey (KiDS)\footnote{\url{http://kids.strw.leidenuniv.nl}.} \cite{2007.15632}, and several experiments are currently in construction, most notably the Rubin Observatory Legacy Survey of Space and Time (LSST)\footnote{\url{https://www.lsst.org}.}, Euclid\footnote{\url{https://www.euclid-ec.org}.} and the Roman Telescope\footnote{\url{https://roman.gsfc.nasa.gov}.}.

    The analysis of photometric survey data is challenging, both in terms of avoiding biases in cosmological parameters from observational systematics, as well as in terms of theoretical modeling of the expected signal. In order to mitigate modeling uncertainties, the analysis of the DES Year 1 galaxy clustering data, alone \citep{1708.01536} or in combination with cosmic shear \citep{1708.01530}, was limited to large scales, where a linear bias relation could be safely used. In this model, it is assumed that the contrast in the number densities of galaxies is proportional to that of the dark matter. While deceptively simple, this model can be shown to be exact in the limit of infinitely large scales under rather general assumptions, as long as the galaxy formation process is local on some finite scale. It is therefore sufficient on large enough scales, but requires that the information on the best measured intermediate and small scales be discarded. The purpose of this paper is to investigate the implications of a re-analysis of the DES Y1 data by employing a more sophisticated bias model.
    
    There are two main difficulties in modeling data on smaller scales. On one hand, the dark matter clustering becomes non-linear, and on the other hand, the manner in which galaxies trace the underlying dark matter fluctuations (which are assumed to be the dynamically dominant component) is also non-linear. The dark matter clustering can be modelled through high-accuracy dark matter $N$-body simulations; this modeling would be exact if it were not for the hydrodynamical effects of baryons, which are luckily confined to scales $k\gtrsim1$ \hMpc, and whose residual effects on large scales can be emulated with numerous approximate schemes \cite{1104.1174,1105.1075,1405.7423,1809.01146,1810.08629}. The tracing of dark matter by galaxies, however, is a more difficult problem. It is hopeless to model the richness of galaxy formation physics completely ab-initio in cosmologically relevant volumes, and even the most precise small-scale hydro simulations need to model subgrid physics phenomenologically \cite{1801.08559}.  Nevertheless, there are numerous phenomenological models, which can be surprisingly successful at explaining galaxy clustering to deeply non-linear scales. Many of these models mostly rely on the notion of halo occupation distribution (HOD) \citep{astro-ph/0206508,astro-ph/0005010}, namely the idea that galaxies occupy dark matter halos with statistics that mostly depend on halo mass. The issue is that, while these are very successful in explaining the observed clustering at 10-percent level, it is hard to set up models that are precise at the percent level without at least some a-posteriori tuning of the model to fit the data. The basic difficulty lies in the fact that these models are under little theoretical control, and are instead based on heuristic models of halo occupation, rather than fundamental principles. Thus there are no strict rules on when to turn left and when to step forward in this dance.

    On the other end of the spectrum of modeling approaches lie purely analytical models that start with a linear-biasing scheme and use perturbative expansion to model both the non-linear dark matter structure growth and the galaxy tracing, integrating any residual small-scale effects into renormalization of the large-scale quantities. This is known as the effective theory of large-scale structure (see e.g. \cite{1406.7843,1611.09787}). While this approach is exact and under full theoretical control, it has the downside that it leads to only a modest gain in the smallest scales that one can still model.
    
    In this paper we employ a method first proposed in \cite{1910.07097} that combines the accuracy of $N$-body simulations with the theoretical robustness of analytical bias expansions. In this approach $N$-body simulations provide accurate information about statistics of dark matter clustering that are required to calculate the galaxy clustering for an analytical bias model in Lagrangian space. In particular, the $N$-body simulations are used to calculate 15 ``basis'' power spectra which multiply the bias coefficients when modeling the galaxy signal. The beauty of this approach is that it maintains the near-exactness of $N$-body simulations for dark matter clustering with the theoretical control of an analytical bias expansion. We refer to this model as the Hybrid EFT (HEFT) model.

    The HEFT method is most  naturally expressed in Fourier space, because any analytic Taylor expansion will break down on sufficiently small Fourier scales. Therefore, on the data side, the main difference with the DES Y1 analysis is that we perform a power spectrum rather than a correlation function analysis. We re-measure the auto- and cross-power spectra and covariance matrix, but otherwise leave the DES analysis largely unchanged, we use the same input catalogs, tomographic bins, and systematic models.
    
    This paper is structured as follows. In Section \ref{sec:data-power-spectrum} we discuss how we measure auto- and cross-power spectra of the DES Y1 data, including estimation of the covariance matrix. This section relies heavily on our previous works using galaxy and shear data from HSC and DES \cite{1912.08209,2010.09717}. In Section \ref{sec:theory} we discuss the theoretical underpinning of the HEFT model, and method used to build an emulator of the HEFT basis power spectra from the \abacus{} simulations. We then proceed to analyze the DES Y1 data using the HEFT model in Section \ref{sec:res}, where we also spend some time addressing variations of the parameter inference to investigate the sensitivity to various analysis choices. We conclude in Section \ref{sec:conclusions} with a recap of our main results and list of potential caveats.
    
  \section{Data and power spectrum measurement}\label{sec:data-power-spectrum}
    \subsection{Data}\label{ssec:data.data}
      We make use of the Dark Energy Survey's first-year (Y1) public data release \cite{2018ApJS..239...18A}. In particular, we use the publicly available key {\tt Y1KP} catalogs\footnote{\url{https://des.ncsa.illinois.edu/releases/y1a1/key-catalogs}}, used to derive cosmological constraints from the joint analysis of galaxy clustering and cosmic shear data (the so-called ``$3\times2$-point'' analysis), presented in \cite{1708.01530} (DY1 hereon). The resulting clustering and shear samples cover an area of $\sim1320\,{\rm deg}^2$.
  
      As a galaxy clustering sample we use the \redmagic{} catalog described in \cite{1708.01536}. The \redmagic{} algorithm selects luminous red galaxies in a way that minimizes photometric redshift uncertainties, producing samples with small redshift distribution tails, well-suited for cosmological galaxy clustering studies. We divide the sample into the same 5 redshift bins used in DY1, and use the same galaxy weights to correct for sky systematics. In the analysis of these data, we make use of the fiducial redshift distributions published with the Y1 dataset, as well as the same model to marginalize over photo-$z$ uncertainties in them (see Section \ref{ssec:params}). Further details can be found in \cite{1708.01536}.
      \begin{figure}
        \centering
        \includegraphics[width=0.8\textwidth]{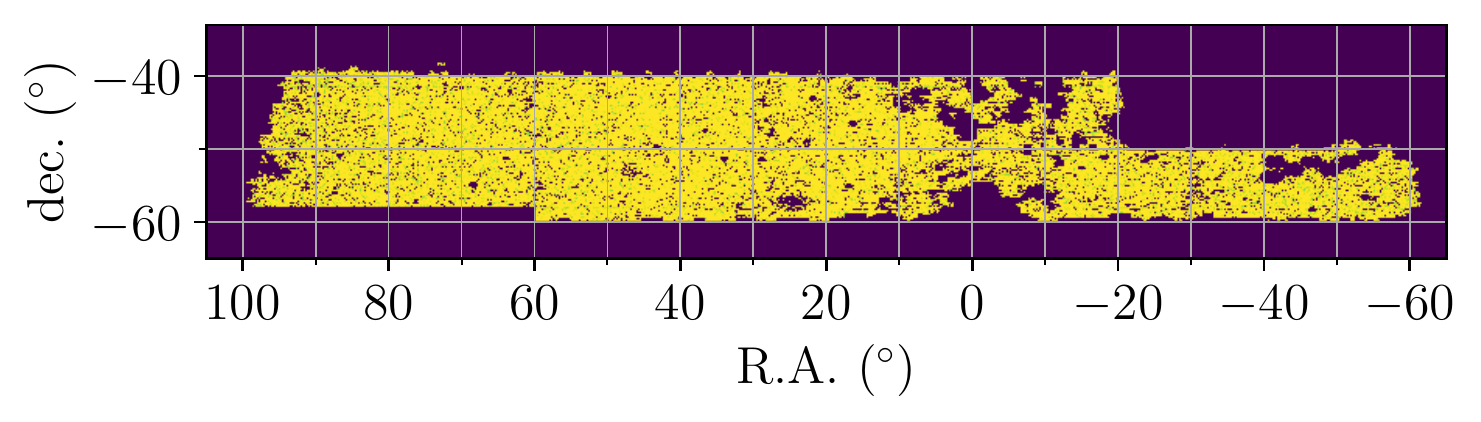}
        \includegraphics[width=0.6\textwidth]{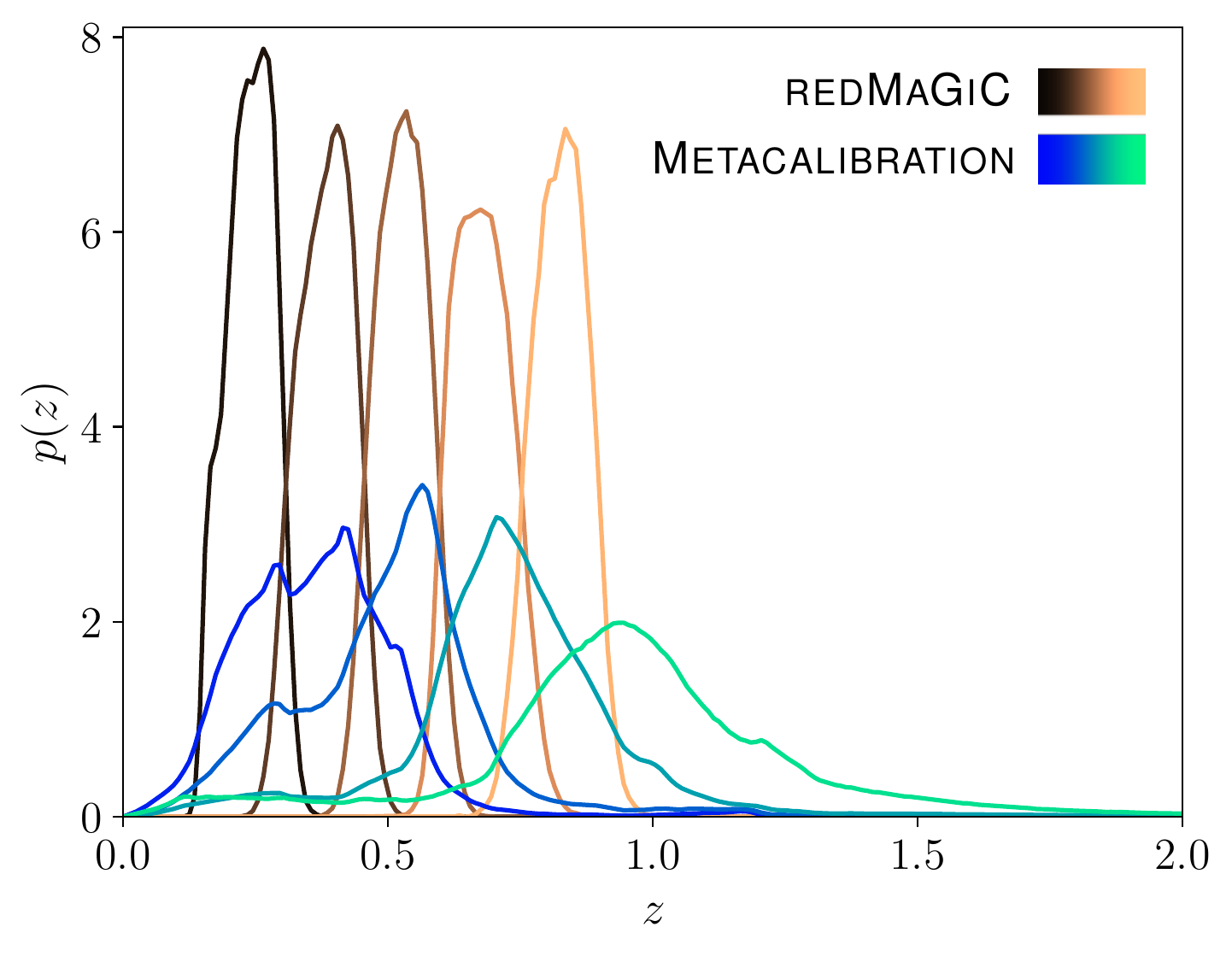}
        \caption{{\sl Top:} one of the angular window functions for the DES Y1 data analysed here. {\sl Bottom:} normalized redshift distribution of the galaxy clustering (\redmagic{}) and cosmic shear (\mcal{}) samples.}
        \label{fig:data}
      \end{figure}
  
      The DES Y1 analysis was carried out using two different shape-measurement algorithms: \mcal{}~\cite{1702.02600,1702.02601} and \textsc{IM3SHAPE}~\cite{1708.01533}. We restrict the current analysis to the \mcal{} sample, applying the same cuts used in \cite{1708.01533}, including its division into four tomographic reshift bins (see also \cite{2010.09717}). We use the redshift distributions provided with the Y1 release for these bins, details for which can be found in \cite{1708.01532}. Details regarding the calibration of galaxy ellipticities, including mean subtraction and the inclusion of selection effects in the \mcal{} response tensor, can be found in \citep{2010.09717}, and follow the same prescriptions used by DY1.

      Fig. \ref{fig:data} shows one of the angular window functions and redshift distributions of the DES Y1 samples used here.

    \subsection{Power spectra and covariances}\label{ssec:data.cls}
      Starting from the official Y1 catalogs described in the previous section, we compute a set of angular auto- and cross-power spectra between the clustering and shear samples. Thus, although our analysis is based on the same samples as DY1, we construct an independent data vector, instead of using the summary statistics (real-space angular correlation functions). We do this in order to take advantage of recent improvements in the algorithms used for the estimation of angular power spectra of LSS data and their covariance \citep{1809.09603,1906.11765,2010.09717}, which allow us to produce robust goodness-of-fit tests, vital for this analysis (see Section \ref{ssec:res.main}). The use of angular power spectra also allows us to impose simple scale cuts that are more directly connected with a comoving Fourier scale $k_{\rm max}$. Finally, our analysis can also validate the results found by the DES Y1 analisis in real space with an independent harmonic-space pipeline (see also \citep{2011.06469}).

      Power spectra were calculated using a pseudo-$C_\ell$ approach \citep{astro-ph/0105302} as implemented in \nmt{}. The details of the pseudo-$C_\ell$ algorithm are described in detail in \cite{1809.09603}, and we provide only a succinct summary here. Pseudo-$C_\ell$ estimators are a fast implementation of the optimal quadratic estimator \citep{astro-ph/9611174}. The algorithm is based on assuming a diagonal map-level covariance matrix, such that inverse-variance weighting is equivalent to a simple multiplication of the sky map by a ``weights map'' or ``mask'' $w(\nv)$. Calculating the power spectrum of two weighted maps is then reduced to averaging the product of their spherical harmonic coefficients over the multipole order $m$, and the time-consuming mode-coupling matrix $M_{\ell\ell'}$ can be computed analytically using fast algorithms to calculate Wigner 3-$j$ symbols. The method works for fields of arbitrary spin, including the galaxy overdensity (spin-0) and cosmic shear (spin-2). We compute all auto- and cross-correlations between the 5 \redmagic{} and 4 \mcal{} redshift bins. Note that, as done in DY1, we discard cross-correlations between different galaxy clustering bins in the likelihood analysis, given their sensitivity to photometric redshift systematics. All maps were generated using \texttt{HEALPix}\footnote{\url{http://healpix.sourceforge.net}.}~\cite{astro-ph/0409513} with resolution parameter $N_{\rm side}=4096$, corresponding to pixels of size $\delta\theta_{\rm pix}\sim1'$. This is small enough that, as described in \cite{2010.09717}, all pixelization effects can be ignored on the scales used in this analysis (see Section \ref{ssec:params}).
  
      The galaxy overdensity maps are estimated using the same method described in \cite{1912.08209}. The overdensity $\delta_p$ at pixel $p$ is computed as $\delta_p=N_p/(w_p\,\bar{N})-1$, where $N_p$ is the weighted number of objects in the pixel, $\bar{N}$ is the mean weighted number of objects per pixel, and $w_p$ is the unmasked fractional area of pixel $p$. The mean is computed as $\bar{N}=\sum_p N_p/\sum_p w_p$. The fractional area map $w_p$, which we also use as the weights map for the galaxy overdensity in the pseudo-$C_\ell$ estimator, is provided in the Y1 data release. The overdensity is set to zero in all fully masked pixels. Furthermore, to avoid noise in strongly masked pixels, we set the weights map to zero for pixels where $w_p<0.5$. Finally, the galaxy noise power spectrum is estimated as described and validated in \cite{1912.08209}, with an extra correction factor \citep{2017MNRAS.467.2085G} to account for effect of galaxy weights. This is then subtracted from all auto-correlations.

      The auto- and cross-correlations involving shear maps only were computed using the procedure described in \cite{2010.09717}. The weights map associated with a shear redshift bin is simply proportional to the sum of shape-measurement weights of all galaxies in the pixel, and the associated shear map is the weighted mean calibrated ellipticity in each pixel. The resulting shear power spectra were presented and validated in \cite{2010.09717}. The clustering-shear cross-correlations were simply estimated as the pseudo-$C_\ell$ between the corresponding galaxy overdensity and shear maps. As a sanity check, we verify that all cross-correlation involving shear $B$-modes are compatible with zero by examining the probability-to-exceed (PTE) of their $\chi^2$ with respect to the null hypothesis.

      All power spectra were calculated for a set of $\ell$-bins (bandpowers) covering the range $\ell\in[0, 12288)$. We use a linear spacing with $\Delta\ell=30$ up to $\ell=240$, and a logarithmic spacing thereafter with $\Delta\log_{10}\ell=0.055$. Even after correcting for the effects of survey geometry by inverting the binned mode-coupling matrix, residual mode-coupling remains as a result of binning. We account for this exactly by convolving the theory prediction with the bandpower window functions as described in \cite{1809.09603}.

      The covariance matrix of these measured power spectra is computed using analytical methods. As discussed in \cite{0810.4170,1302.6994}, the main contributions to the covariance of power spectra of large-scale structure tracers can be written as:
      \begin{equation}
        {\rm Cov}_{\ell\ell'}={\rm Cov}^{\rm G}_{\ell\ell'}+{\rm Cov}^{\rm cNG}_{\ell\ell'}+{\rm Cov}^{\rm SSC}_{\ell\ell'},
        \label{eq:cov}
      \end{equation}
      where ${\rm Cov}^{\rm G}_{\ell\ell'}$ is the ``Gaussian'' covariance matrix, associated with the fields' disconnected trispectrum, and ${\rm Cov}^{\rm cNG}_{\ell\ell'}$ and ${\rm Cov}^{\rm SSC}_{\ell\ell'}$ are non-Gaussian terms, sourced by the non-linear evolution of the matter overdensities under gravity.

      ${\rm Cov}^{\rm G}_{\ell\ell'}$ dominates the error budget, and therefore must be carefully calculated, accounting for the effects of survey geometry in the form of mode-coupling. To do so, we follow the approximate methods of \cite{astro-ph/0307515,1906.11765,2010.09717}. The exact calculation of the covariance matrix scales as $O(\ell_{\rm max}^6)$, and is therefore unfeasible for the range of scales used in this work. The calculation can be reduced to $O(\ell_{\rm max}^3)$ under the approximation of a narrow mode-coupling kernel and a sufficiently flat underlying power spectrum. While this is a good approximation for galaxy clustering, the large inhomogeneity of the weak lensing mask (effectively proportional to the galaxy density) breaks these assumptions and can lead to $O(1)$ errors in the cosmic shear covariance. To remedy this, we make use of the improved narrow-kernel approximation presented in \cite{2010.09717}, which is able to accurately recover the true power spectrum uncertainties up to a few percent on the scales used here, including the different noise and signal contributions.

      We estimate the SSC and cNG contributions to the total covariance, following the halo model based approach of Ref.~\cite{1601.05779}, as was done in Refs.~\cite{1912.08209, 2006.00008, 2010.09717}. For a more detailed description, we refer the reader to these works and the references therein. As the total covariance matrix given in Eq.~\ref{eq:cov} is dominated by the Gaussian part, we model finite sky effects for the non-Gaussian corrections using the approximations given in Ref.~\cite{1601.05779} and do not fully account for mode-coupling as we do for the Gaussian part. Note that, within the range of scales used here, the effect of the non-Gaussian terms on the $\chi^2$ is smaller than 2\%, and therefore this should be a good approximation as long as survey geometry effects are accurately accounted for in the Gaussian part.

  \section{Modeling the signal}\label{sec:theory}\label{sec:modeling-signal}
  \subsection{Projected statistics}\label{ssec:theory.cls}
      We will extract constraints on cosmological parameters from the two-point statistics of two fields projected on the celestial sphere: the galaxy overdensity $\delta_g^\alpha(\nv)$ and the weak lensing shear $\gamma^\alpha(\nv)$ for galaxies in redshift bin $\alpha$. These are related to the three-dimensional fluctuations in the galaxy number density $\Delta_g({\bf x})$ and the matter density $\Delta_m({\bf x})$ via \citep{astro-ph/9912508,1706.09359}
      \begin{align}\nonumber
        &\delta_g^\alpha(\nv)=\int_0^{\chi_H}d\chi\,q^\alpha_g(\chi)\,\Delta_g(\chi(z)\nv, z),\hspace{12pt}
        \gamma^\alpha(\nv)=\int _0^{\chi_H}d\chi\,q^\alpha_\gamma(\chi)\,\left[-\chi^{-2}\eth\eth\nabla^{-2}\Delta_m(\chi\nv,z)\right],\\\label{eq:deltagamma}
        &q^\alpha_g(\chi)\equiv\frac{H(z)}{c}p_\alpha(z),\hspace{12pt}q^\alpha_\gamma(\chi)\equiv\frac{3}{2}H_0^2\Omega_m\frac{\chi}{a(\chi)}\int_{z(\chi)}^\infty dz' p_\alpha(z')\frac{\chi(z')-\chi}{\chi(z')},
      \end{align}
      where $c$ is the speed of light, $\nv$ is the sky direction, $\chi$ is the comoving radial distance at redshift $z$, $\chi_H$ is the distance to the horizon, $H(z)$ is the Hubble expansion rate, $H_0\equiv H(z=0)$, $\Omega_m$ is the matter density parameter today, $p_\alpha(z)$ is the redshift distribution in bin $\alpha$, and $\eth$ is the spin-raising differential operator, acting on a spin-$s$ quantity as:
      \begin{equation}
        \eth\,_sf(\theta,\varphi)=-(\sin\theta)^s\left(\frac{\partial}{\partial\theta}+\frac{i}{\sin\theta}\frac{\partial}{\partial\varphi}\right)(\sin\theta)^{-s}\,_sf
      \end{equation}
      and turning it into a spin-$(s+1)$ quantity.

      The power spectrum between quantities $X$ and $Y$ ($\delta_g$ or $\gamma$) in bins $\alpha$ and $\beta$ respectively, $C_\ell^{(X,\alpha)(Y,\beta)}$ is the covariance of the spherical harmonic coefficients of both fields, and can be related to the power spectrum of the three-dimensional quantities associated with $X$ and $Y$ ($\Delta_g$ or $\Delta_M$) $P_{XY}(k,z)$ via:
      \begin{equation}\label{eq:limber}
        C_\ell^{(X,\alpha),(Y,\beta)}=\int\frac{d\chi}{\chi^2}\,q_X^\alpha(\chi)\,q_Y^\beta(\chi)\,P_{XY}\left(k=\frac{\ell+1/2}{\chi},z(\chi)\right).
      \end{equation}

      Equation \ref{eq:limber} uses Limber's approximation \citep{1953ApJ...117..134L,astro-ph/0308260}, valid for the wide radial kernels considered here. In order to account for the difference between angular and three-dimensional derivatives in Eq. \ref{eq:deltagamma} (i.e. $\chi^2\eth^2\nabla^{-2}\not\equiv1$), the lensing kernel must be multiplied by an $\ell$-dependent prefactor
      \begin{equation}
        G_\ell\equiv\sqrt{\frac{(\ell+2)!}{(\ell-2)!}}\frac{1}{(\ell+1/2)^2},
      \end{equation}
      which becomes irrelevant (sub-percent) for $\ell>11$ \citep{1702.05301}.
  
      Thus, given a set of cosmological parameters, and the redshift distributions of all bins considered, all that remains to specify is the 3-dimensional power spectra between $\Delta_m$ and $\Delta_g$. For the matter power spectrum $P_{mm}(k,z)$, we use the non-linear prediction from \texttt{HALOFIT} \cite{astro-ph/0207664,1208.2701}, as was done in DY1. We describe the model used to describe the galaxy-matter connection in the next section.

    \subsection{Hybrid EFT model for galaxy clustering}\label{ssec:theory.heft}
      We follow a perturbative effective field theory (EFT) approach to galaxy biasing in Lagrangian space, coupled with the non-linear dynamical evolution of $N$-body simulations as prescribed in \cite{1910.07097}. We describe the logic behind this Hybrid EFT method here, and refer readers to \cite{1910.07097,1611.09787,0807.1733} for a detailed description of galaxy bias.

      The complex physical processes that govern the formation and evolution of galaxies necessarily imply a complex relationship between their distribution and that of the matter inhomogeneities they trace. In general, this relation should be non-linear, non-local and, without exact knowledge of the small-scale physics, stochastic. Non-local and stochastic effects are sourced by the dependence of the galaxy abundance at a given point in space on the small-scale physical processes in a region around it, and therefore should become negligible on scales larger than the size of this region (e.g. the Lagrangian radius of a typical dark matter halo).

      On scales where galaxy formation can be modelled as a local process, it is then possible to invoke the equivalence principle, in terms of which the leading gravitational effects are associated with the Hessian of the gravitational potential $\partial_i\partial_j\Phi$. This can be split into its scalar trace, proportional to the matter overdensity $\delta$, and the traceless tidal tensor $s_{ij}\equiv(\partial_i\partial_j\nabla^{-2}-1/3)\delta$. On these scales, we can therefore describe the number overdensity of galaxies found at redshift $z$ in the Lagrangian initial conditions as a general functional of the local $\delta$ and $s_{ij}$. Expanding this functional up to second order:
      \begin{equation}\label{eq:lbexp}
        1+\Delta_{g,L}=1+b_1\delta_L+b_2(\delta_L^2-\langle\delta_L^2\rangle)+b_s(s^2_L-\langle s^2_L\rangle)+b_{\nabla}\nabla^2\delta_L,
      \end{equation}
      where we have only kept scalar combinations of $s_{ij}$, $s^2\equiv s_{ij}s^{ij}$, and we have included a leading-order non-local contribution $\propto\nabla^2\delta_L$. The subscript $L$ is a reminder that all quantities are evaluated at the initial Lagrangian coordinates ${\bf q}$. One can then evolve $\Delta_{g,L}$ to its observed redshift by advecting the Lagrangian $\Delta_g$ to the final Eulerian coordinates:
      \begin{equation}\label{eq:lpt}
        1+\Delta_g({\bf x})=\int d^3{\bf q}\,[1+\Delta_{g,L}({\bf q})]\,\delta^D({\bf x}-{\bf q}-\Psi({\bf q})),
      \end{equation}
      where $\Psi$ is the Lagrangian displacement vector.

      This calculation can be done using Lagrangian perturbation theory (e.g. \cite{2012.04636}) or, as proposed in \cite{1910.07097}, solving the full non-linear evolution in an $N$-body simulation. Substituting Eq. \ref{eq:lbexp} into \ref{eq:lpt}, the final galaxy overdensity is a linear combination of the individual operators in Eq. \ref{eq:lbexp} ($\delta_L$, $\delta_L^2$, $s^2_L$ and $\nabla^2\delta_L$) advected to the Eulerian positions. These fields can be calculated in the simulation by weighting each matter particle in a given snapshot by the value of the corresponding operator in the initial conditions at the original Lagrangian coordinates. The cross-power spectrum between the galaxy and matter overdensities, as well as the galaxy-galaxy power spectrum are then given by:
      \begin{equation}\label{eq:pkexp}
        P_{gm}(k)=\sum_{\alpha\in{\cal O}}b_\alpha P_{1\alpha}(k),\hspace{12pt}
        P_{gg}(k)=\sum_{\alpha\in{\cal O}}\sum_{\beta\in{\cal O}}b_\alpha b_\beta P_{\alpha\beta}(k)
      \end{equation}
      where ${\cal O}\equiv\{1,\delta_L,\delta_L^2,s_L^2,\nabla^2\delta_L\}$ is the full set of second-order operators in Eq. \ref{eq:lbexp}, $P_{\alpha\beta}(k)$ is the power spectrum of the advected fields $\alpha$ and $\beta$, and $b_\alpha$ are the corresponding bias coefficients. In this formalism $P_{11}(k)$ is the non-linear matter power spectrum, and its corresponding bias parameter (called $b_0$ here) is $b_0=1$. Our bias model then reduces to computing a set of 15 different power spectrum templates $P_{\alpha\beta}(k)$ for the five different operators. We construct these from the \abacus{} suite of simulations, described in the next section.

      We note that while formally Lagrangian and Eulerian models are both valid and complete descriptions, their predictions do not match at any given order. Therefore, even the simplest linear analysis will give different results in Eulerian and Lagrangian spaces for a finite maximum wavenumber $k_{\rm max}$.
  
      Finally, it is also common to include an additive stochastic term $\epsilon$ in the bias expansion (Eq. \ref{eq:lpt}), to account for the impact of small-scale density fluctuations on galaxy formation. In the simplest case, this effect can be modelled by treating galaxies as a Poisson sampling of an underlying smooth galaxy density field, in which case the contribution to the galaxy auto-correlation is $P_{\epsilon\epsilon}=1/\bar{n}_g$, where $\bar{n}_g$ is the comoving number density of galaxies. In our analysis, we will also consider the effect of departures from this pure shot-noise contribution, of the form $P_{\epsilon\epsilon}=A_\epsilon/\bar{n}_g$, with $A_\epsilon\neq1$ (see Section \ref{ssec:params} and Section \ref{ssec:res.tests}).

    \subsection{Abacus simulations}\label{ssec:theory.abacus}
      To build a model for the different power spectrum templates $P_{\alpha\beta}(k)$ in Eq. \ref{eq:pkexp} we use the \abacus{} suite of $N$-body simulations \cite{1712.05768,Maksimova:2021xxx,Garrison:2021xxx}.
      
      \abacus{} was designed to meet the cosmological simulation requirements of the Dark Energy Spectroscopic Instrument (DESI) survey and run on the Summit supercomputer at the Oak Ridge Leadership Computing Facility. The simulations are run with the highly accurate \textsc{Abacus} cosmological $N$-body simulation code \citep{2019MNRAS.485.3370G}, optimized for GPU architectures and large-volume, moderately clustered simulations. The \textsc{Abacus} code is extremely fast, performing 70 million particle updates per second on each of the Summit nodes, and also extremely accurate, with typical force accuracy below $10^{-5}$. The output halo catalogs and particle subsamples (\texttt{A} and \texttt{B}, amounting to 10\% of the total particle population) are organized into 12 primary redshift snapshots ($z = $ 0.1, 0.2, 0.3, 0.4, 0.5, 0.8, 1.1, 1.4, 1.7, 2.0, 2.5, and 3.0). The data products have been designed with the aim of supporting mock catalogs to be constructed using halo occupation distributions, as well as efficient access to measurements of the density fields. In this work, we use the 7 redshift snapshots at $z \leq 1.1$, $z = 0.1$, 0.2, 0.3, 0.4, 0.5, 0.8, 1.1, covering the redshift range of the DES Y1 clustering sample, and the particle subsample \texttt{A}, which contains 3\% of the total particle content.
      \begin{table}
        \begin{center}
          \begin{tabular}{| c | c | c | c | c |}
          \hline\hline
          Sim. name & $\Omega_b h^2$ & $\Omega_c h^2$ & $n_s$ & $\sigma_8$ \\ [0.5ex]
          \hline
          \texttt{c000} & 0.02237 & 0.1200 & 0.9649 & 0.811355 \\ [1ex]
          \texttt{c100}$/$\texttt{c101} & 0.02282$/$0.02193 & -- & -- & -- \\ [1ex]
          \texttt{c102}$/$\texttt{c103} & -- & 0.1240$/$0.1161 & -- & -- \\ [1ex]
          \texttt{c104}$/$\texttt{c105} & -- & -- & 0.9749$/$0.9549 & -- \\ [1ex]
          \texttt{c112}$/$\texttt{c113} & -- & -- & -- & 0.8276$/$0.7954 \\ [1ex]
          \hline
          \hline
          \end{tabular}
        \end{center}
        \caption{Simulations from the \abacus{} suite used to generate the power spectrum templates used in this analysis.}\label{tab:sims}
      \end{table}

      The \abacus{} suite contains simulations for various cosmologies. Here, we employ the fiducial simulation \texttt{AbacusSummit\_base\_c000\_ph000} at base resolution (6912$^3$ dark matter particles in a box of length 2000 Mpc$/h$), and some of the ``linear derivatives'' simulations at the same resolution. These are used to account for the parameter dependence of the power spectrum templates as described below. The simulations used are listed in Table \ref{tab:sims}. The parameters that define the fiducial cosmology are the baryon energy density ($\Omega_b h^2 = 0.02237$), the cold dark matter energy density ($\Omega_c h^2 = 0.1200$), the primordial tilt ($n_s = 0.9649$), the amplitude of matter fluctuations ($\sigma_8 = 0.811355$), and the distance to last scatter ($100 \theta_\ast = 1.041533$). The ``linear derivative'' simulations vary each of these in turn, as indicated by the columns in the table.

      For each simulation and redshift bin we produce 15 power spectrum templates, corresponding to the auto- and cross-correlation between the 5 operators of the Lagrangian bias expansion (Eq. \ref{eq:lpt}) following the method outlined in \cite{1910.07097}:
      \begin{enumerate}
        \item We use the initial conditions to calculate 4 fields: $\delta_L$, $\delta_L^2$, $s^2_L$ and $\nabla^2\delta_L$. These fields are computed on a cubic grid of size $2304^3$. Following \citep{1910.07097,2101.11014} we do not apply any smoothing and so our results must depend weakly on cell size.  
        \item At each snapshot, we evolve the initial condition fields to the corresponding redshift assuming a linear growth factor. Then, for each of the 5 bias operators ($1,\,\delta_L,\,\delta_L^2,\,s_L^2,\,\nabla^2\delta_L$), we compute their advected version by assigning each dark matter particle a weight given by the value of the corresponding operator at the particle's position in the initial conditions.
        \item Finally, we compute and store the auto- and cross-power spectra between all advected fields.
      \end{enumerate}

\begin{figure}
        \centering  
        \includegraphics[width=1.\textwidth]{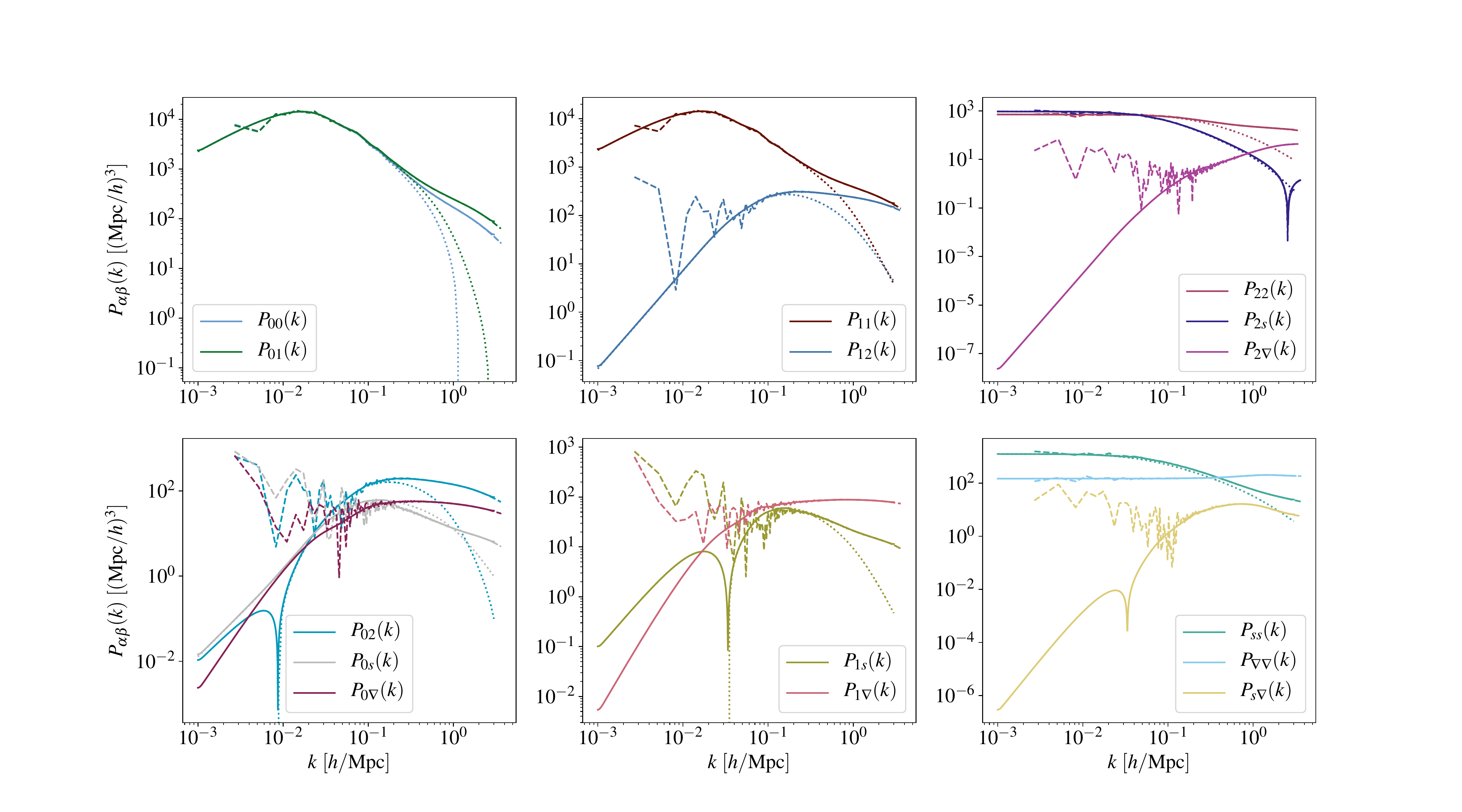}
        \caption{Power spectrum templates at $z = 0.5$ for the fiducial \abacus{} simulation (\texttt{c000}). The dotted lines correspond to the 1-loop LPT power spectra computed using  \texttt{velocileptors} \citep{2005.00523}, while the dashed lines are derived from \abacus{}. In solid lines, we show the combined power spectra, obtained by smoothly interpolating between both. As expected, the 1-loop LPT theory lacks small-scale power compared with the $N$-body result. We note that 1-loop LPT makes no prediction for the templates involving cross-correlations with the $\nabla^2 \delta_L$ field, for which we assume $P_{X \nabla}(k) \approx k^2 P_{X \delta}$ on large scales and the numerical values on small scales. We model the $P_{\nabla \nabla}(k)$ as approximately constant on large scales.}\label{fig:templates}
      \end{figure}

      The power spectrum templates thus calculated are noisy on large scales due to cosmic variance. To avoid this, we combine the simulated power spectra with theoretical predictions from Lagrangian perturbation theory (LPT) using the \texttt{velocileptors} code \cite{2012.04636,2005.00523} on large scales.  We use the following combination of both predictions to enforce a smooth transition between them at $k\sim k_{\rm pivot}$:
      \begin{equation}
        P_{\alpha\beta}(k)=(1-w(k))P^{\rm sim}_{\alpha\beta}(k)+w(k)\,P^{\rm LPT}_{\alpha\beta}(k),
      \end{equation}
      where $P^{\rm sim}_{\alpha\beta}$ and $P^{\rm LPT}_{\alpha\beta}$ are the predictions from \abacus{} and LPT respectively, and weighting function
      \begin{equation}
        w(k)\equiv \frac{1}{2}\left[1-{\rm tanh}\left(\frac{k-k_{\rm pivot}}{\Delta k_w}\right)\right],
      \end{equation}
which ensures smooth interpolation between the two limits. We use $\Delta k_w=0.01\,h{\rm Mpc}^{-1}$, but manually fine-tune values of $k_{\rm pivot}$ for the different operator combinations, based on their large-scale behavior:
      \begin{align}
        &k_{\rm pivot}^{0,s}=k_{\rm pivot}^{1,s}=k_{\rm pivot}^{\nabla,s}=0.2\,h{\rm Mpc}^{-1},\hspace{12pt}
        k_{\rm pivot}^{2,s}=k_{\rm pivot}^{s,s}=0.03\,h{\rm Mpc}^{-1},\\
        &k_{\rm pivot}^{2,\nabla}=0.3\,h{\rm Mpc}^{-1},\hspace{12pt}
        k_{\rm pivot}^{1,\nabla}=0.07\,h{\rm Mpc}^{-1},\hspace{12pt}k_{\rm pivot}^{\rm other}=0.09\,h{\rm Mpc}^{-1}.
      \end{align}
      The power spectrum templates thus produced for the fiducial \texttt{c000} simulation at $z=0.5$ are shown in Fig. \ref{fig:templates}, which also illustrates the procedure outlined above. The code used to generate these templates is available at \url{https://github.com/boryanah/hybrid_eft_nbody}.
  
      In order to account for the dependence of the power spectrum templates on cosmological parameters, we use a linear Taylor expansion around the fiducial \abacus{} cosmology (first row of Table \ref{tab:sims}). We first compute an estimate of the derivative of the power spectrum templates with respect to the four cosmological parameters $(\Omega_bh^2,\Omega_ch^2,\sigma_8,n_s)$ via numerical differentiation of the templates found for the fiducial \texttt{c000} simulation and the linear derivative simulations. I.e. for a given parameter $\theta$:
      \begin{equation}
        \partial_\theta P_{\alpha\beta}(k)=\frac{P_{\alpha\beta}(k;\theta_F+\delta\theta)-P_{\alpha\beta}(k;\theta_F-\delta\theta)}{2\delta\theta},
      \end{equation}
      where  $P_{\alpha\beta}(k;\theta_F+\delta\theta)$ are the templates calculated in the two linear-derivative simulations for which this parameter is varied (by an amount $\delta\theta$), and $\theta_F$ is the value of the parameter in the fiducial \abacus{} cosmology. The power spectrum template at a set of cosmological parameters $\vec{\theta}$ is then given by:
      \begin{equation}\label{eq:pktaylor}
        P_{\alpha\beta}(k;\vec{\theta})= P_{\alpha\beta}(k;\vec{\theta}_*)+(\vec{\theta}-\vec{\theta}_*)\cdot\nabla_\theta P_{\alpha\beta}(k),
      \end{equation}
      where $\vec{\theta}_*$ is the fiducial \abacus{} cosmology.

      Note that the initial conditions of all simulations we use are produced for the same phase (\texttt{ph000}). Thus, when taking finite differences, we eliminate some of the cosmic variance noise. We verified the validity of the linear Taylor expansion by comparing the prediction from Eq. \ref{eq:pktaylor} with power spectrum templates calculated directly from the extended grid of ``linear derivative'' \abacus{} simulations at redshifts $0.5 \leq z \leq 1.1$ (labelled \texttt{c117-120}, \texttt{c119, c120}). The prediction is found to be accurate at the $\sim1\%$ level on all scales of interest $k \leq 1$ \hMpc.

      For parameter values sufficiently far away from the fiducial \abacus{} cosmology, we expect the linear Taylor expansion to break down. Thus, in order to recover some of the true parameter dependence, we combine \abacus{} predictions for the \emph{ratio} between $P_{\alpha\beta}(k)$ and the matter power spectrum $P_{mm}(k)\equiv P_{11}(k)$, with the \texttt{HALOFIT} prediction for the latter as follows: 
      \begin{equation}
        P_{\alpha\beta}(k;\vec{\theta})=\frac{P^{\tt HF}_{mm}(k;\vec{\theta})}{P^{\tt AB}_{11}(k;\vec{\theta})} P^{\tt AB}_{\alpha\beta}(k;\vec{\theta}),
      \end{equation}
      where $P^{\tt HF}$ and $P^{\tt AB}$ are the \texttt{HALOFIT} and \abacus{} predictions respectively. By taking the ratio of the \abacus{} predictions we thus mitigate the impact of residual noise on the templates, and some of the error made in assuming a linear dependence on cosmological parameters around the \abacus{} fiducial model. In particular, this approach recovers the \texttt{HALOFIT} matter power spectrum exactly.

      An important caveat must be noted. The linear derivative simulations available in the \abacus{} suite use a value of the Hubble parameter $h$ determined by holding the comoving angular diameter distance to the last-scattering surface, $\theta_\ast$, constant, and equal to the value inferred by \planck{} \cite{1807.06209}. Effectively, this means that our parametrization of the cosmological dependence of $P_{\alpha\beta}$ is ``missing'' the Hubble parameter, which is assumed to reproduce the position the peak position in the \planck{} CMB power spectrum. This is one of the best and most robustly measured quantities in cosmology; however, it complicates the direct comparison between our results and those of DY1. Although we find that our results are not strongly sensitive to this (see Section \ref{sec:res}), a robust implementation of this method should allow for variation of all basic cosmological parameters.

      As stated above, our ability to account for the dependendence of the power spectrum templates on cosmological parameters accurately is limited by the range of cosmologies covered by the \abacus{} suite. Although combining the simulation data with \texttt{HALOFIT} should allow us to capture some of the dependence beyond the linear Taylor expansion, the current model could clearly be improved by building a more complete power spectrum emulator covering a wider region of the cosmological parameter space. This has recently been done in \cite{2021arXiv210111014K} and \cite{2021arXiv210112187Z}. Nevertheless, as we show in Section \ref{sec:res} by switching off the cosmological dependence of the power spectrum templates completely, our final constraints on $(S_8,\Omega_m)$ are insensitive to the inaccuracy of the linear expansion for the data used here.

\section{Results}\label{sec:res}
\label{sec:results}

    \begin{figure}[ht]
      \centering  
      \includegraphics[width=1.\textwidth]{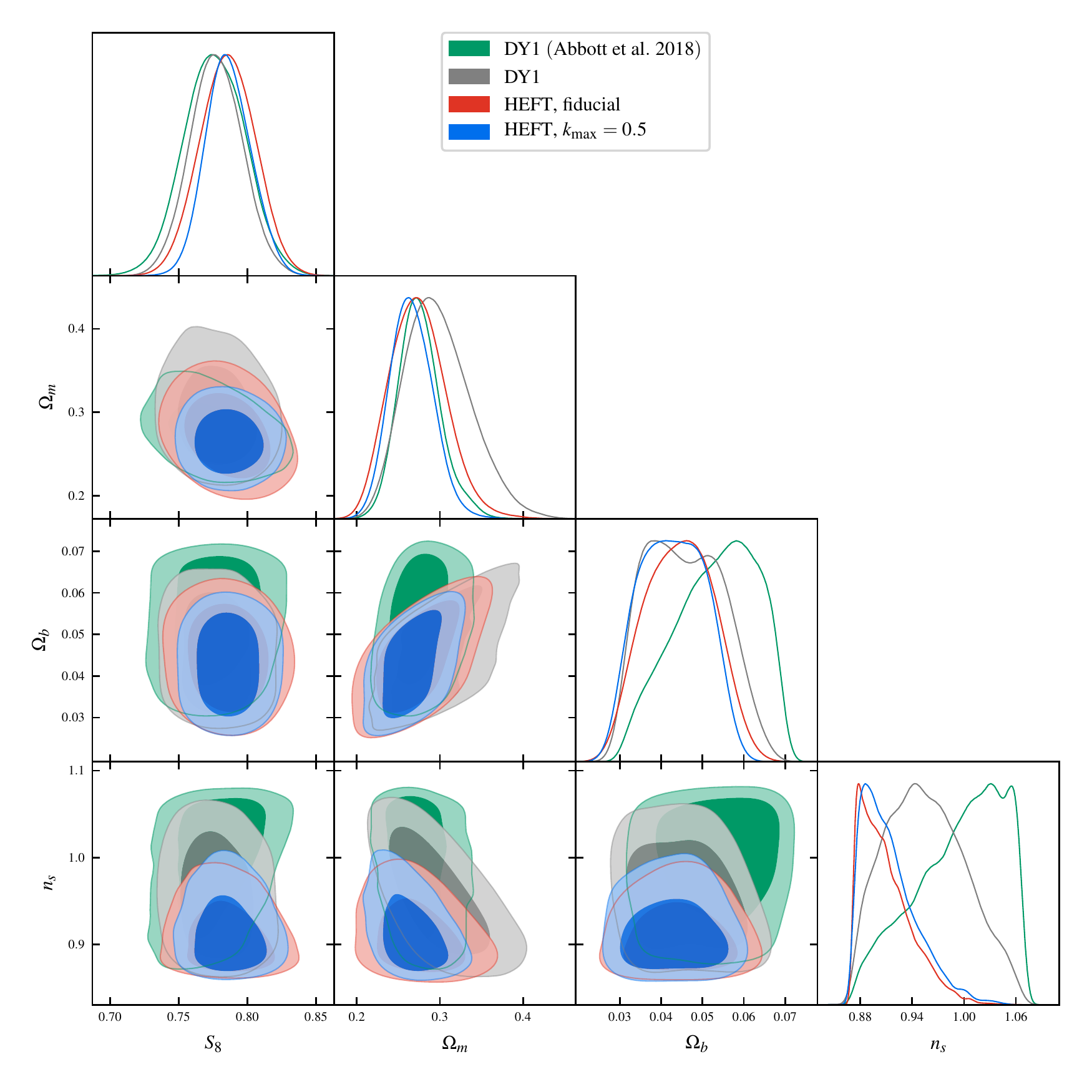}
      \caption{Contour plots showing the constraints on cosmological parameters using a linear bias model, also adopted in the DES Y1 analysis, in gray and the hybrid model presented in this work with $k_{\rm max} = 0.3$ and 0.5 \iMpc{} in red and blue, respectively. In green, we also show the parameter constraints obtained by the DES collaboration in their $3 \times 2$-point analysis \cite{1708.01530}. We see that the hybrid approach is able to place tighter constraints on $\Omega_m$ and $S_8$, while $n_s$ and $\Omega_b$ remain largely unconstrained, in agreement with \cite{1708.01530} (see Section \ref{ssec:res.tests} for a more detailed discussion about $n_s$). The DY1 linear bias constraints on $\Omega_m$ and $S_8$ are consistent with each other and exhibit certain differences attributable to the following: our analysis is performed in harmonic space; it uses a different cosmological parametrization, and additionally does not vary massive neutrino energy density. For a more quantitative assessment, see Table~\ref{tab:constraints}.}\label{fig:nested_0}
    \end{figure}

In this section, we present the cosmological constraints extracted from the DES Y1 galaxy clustering and cosmic shear data for different choices of bias parametrization and scale cuts.

  \subsection{Likelihood analysis}\label{ssec:params}
    \begin{table}
      \begin{center}
      \begin{tabular}{| c  c | c  c |}
        \hline\hline
        Parameter & Prior & Parameter & Prior \\ [0.5ex]
        \hline
        $\Omega_ch^2$ & [0.08, 0.16] & $A_\epsilon$ & ${\cal N}(1,0.1)$\\
        $\Omega_bh^2$ & [0.013, 0.031] & $A_{\rm IA}$ & [-5, 5] \\
        $\sigma_8$ & Free & $\eta_{\rm IA}$ & [-5, 5] \\
        $n_s$ & [0.87, 1.07] &  $m_i$ & Table I in \cite{1708.01530}\\
        $100 \theta_\ast$ & 1.041533 & $\Delta z_g^i$ & Table I in \cite{1708.01530} \\
        $b_0^i,b_1^i,\,b_2^i,\,b_s^i,\,b_\nabla^i$ & [-5, 5] & $\Delta z_s^i$ & Table I in \cite{1708.01530} \\
        \hline
        \hline
      \end{tabular}
      \end{center}
      \caption{Model parameters and priors. An index $i$ denotes parameters with independent copies in each galaxy clustering or shear redshift bin. The definition of the nuisance parameters $(A_{\rm IA},\eta_{\rm IA},m_i,\Delta z_g^i,\Delta_s^i)$ can be found in DY1 \citep{1708.01530}. Square brackets denote a flat prior, whereas ${\cal N}$ denotes a Gaussian prior. In addition to $A_\epsilon$, the latter is adopted also for $m_i,\Delta z_g^i,\Delta_s^i$. The comoving angular distance to last scatter, $ \theta_\ast$, is held fixed in the fiducial case and loosened in one of our tests (see Table~\ref{tab:constraints}).}\label{tab:params}
    \end{table}
    In order to derive constraints on cosmological and bias parameters we use a Gaussian likelihood of the form:
    \begin{equation}
      \log p(\vec{\theta}|{\bf d})=-\frac{1}{2}[{\bf d}-{\bf t}(\vec{\theta})]^T{\sf C}^{-1}[{\bf d}-{\bf t}(\vec{\theta})]+\log p_p(\vec{\theta})+K
    \end{equation}
    where $\vec{\theta}$ is the set of parameters to be constrained, ${\bf d}$ is a data vector of power spectra, ${\sf C}$ is its covariance matrix, ${\bf t}(\vec{\theta})$ is the theory prediction, $p_p(\vec{\theta})$ is the prior distribution, and $K$ is a normalization constant. We sample this likelihood with a modified version of the MCMC sampler \texttt{MontePython}\footnote{\url{https://github.com/boryanah/montepython\_public}}
    \citep{2019PDU....24..260B}, using the Core Cosmology Library \citep{1812.05995} to calculate all angular power spectra contributing to ${\bf t}(\vec{\theta})$.

    In our fiducial case, ${\bf d}$ contains all galaxy auto-correlations, and all galaxy-shear and shear-shear correlations. We impose the following scale cuts: in all cases, shear-shear power spectra are used on scales $\ell<2000$, to avoid the impact of baryonic effects \citep{1809.09148}. Galaxy-galaxy and galaxy-shear correlations, on the other hand, are cut on $\ell<k_{\rm max} \bar{\chi}$, where $\bar{\chi}$ is the distance to the mean redshift of the corresponding galaxy clustering sample, and $k_{\rm max}$ is a comoving cutoff scale. The fiducial value of $k_{\rm max}$ is $0.3$ \iMpc, but we will consider values in the range $[0.15, 0.6]$ \iMpc.

    Our model is described by a number of cosmological, bias and nuisance parameters. We consider variations in five cosmological parameters: the cold dark matter and baryon densities $(\Omega_ch^2,\Omega_bh^2)$, the primordial tilt $n_s$, the amplitude of density fluctuations $\sigma_8$, and the distance to the surface of last scattering $\theta_\ast$. In the fiducial case we fix the latter to the value used in the \abacus{} $100\, \theta_\ast=1.041533$, but we also explore the impact of freeing this parameter. 

    Our fiducial bias model is defined by the set of EFT bias parameters $(b_1,b_2,b_s,b_\nabla)$. We assign different bias parameters for the 5 different galaxy redshift bins, for a total of 20 free parameters. In order to reproduce the linear bias model used in DY1, we also consider a single free bias parameter $b_0$ per redshift bin, while keeping all other bias parameter fixed to zero (note that $b_0=1$ in the HEFT model). When reproducing the DY1 analysis we used a scale cut $k_{\rm max}=0.15$ \iMpc, roughly corresponding to the inverse of the minimum comoving separation used in DY1. As an extension to the 4-parameter HEFT model, we also marginalized over a stochastic bias parameter with a flat power spectrum. We do so by scaling the shot noise power spectrum with a free amplitude parameter $A_\epsilon$ with a $10\%$ Gaussian prior centered on $A_\epsilon=1$. The width of the prior was determined by comparing the noise power spectrum estimated as described in Section \ref{ssec:data.cls} with the galaxy power spectrum at the high-$\ell$, noise-dominated regime.

    Apart from the bias and cosmological parameters, we also vary 15 nuisance parameters, describing other sources of systematic uncertainty in the DES Y1 data. These include shifts in the mean of the redshift distribution in each tomographic bin, multiplicative shear bias parameters, and a two-parameter intrinsic alignment model. The details are described in the DY1 paper \cite{1708.01530}, and we use the same priors specified there. The full set of model parameters and priors used are listed in Table \ref{tab:params}.

  \subsection{Comparison between the linear and HEFT bias models}\label{ssec:res.main}
    One of our main goals is comparing the final constraints achievable through a linear bias model in a reduced range of scales (as done e.g. in DY1), with those found using the HEFT model using data on smaller, mildly non-linear scales. For concreteness, the parameter space in both cases is defined as follows:
    \begin{itemize}
      \item DY1: we vary a single linear bias parameter $b_0$ in each galaxy clustering redshift bin. We also vary four cosmological parameters ($\Omega_ch^2,\,\Omega_bh^2,\,n_s,\,\sigma_8$), and 15 nuisance parameters, for a total of 24 free parameters.
      \item HEFT: we vary four bias parameters ($b_1$, $b_2$, $b_s$, $b_\nabla$) in each redshift bin, while keeping $b_0=1$, leading to a 39-dimensional parameter space when combined with the cosmological and nuisance parameters.
    \end{itemize}
    The linear bias model will be restricted to comoving scales $k_{\rm max}\lesssim0.15$ \iMpc, while we will present results for the HEFT case as a function of $k_{\rm max}$. Note that, although we label it ``DY1'', the linear bias model is slightly different from that used by \cite{1708.01530} since, while the Hubble constant was a free parameter in their analysis, in the fiducial case we determine it by holding $\theta_\ast$ fixed.

    We will quantify the goodness of fit of a given bias model in terms of the probability-to-exceed (PTE) of the model's minimum $\chi^2$ value. The PTE depends on the total number of degrees of freedom $\nu$, which in turn depends on the number of free parameters in the model. For a linear model with $N_\theta$ linearly independent free parameters with unconstrained priors, $\nu$ would be simply $N_{\rm data}-N_\theta$. In the presence of non-linear parameters and tight priors, the definition of $\nu$ is less clear. Here we use an effective number of degrees of freedom $\nu_{\rm eff}$ determined as follows. We generate a synthetic data vector drawn from a multi-variate Gaussian distribution with a mean given by the theoretical prediction for a set of fiducial parameters $\theta_{\rm fid}$, and the power spectrum covariance described in Section \ref{ssec:data.cls}. We then run a $\chi^2$ minimizer varying the cosmological, bias and nuisance parameters, to find the best fit parameters $\theta_{\rm bf}$, and compute the difference in $\chi^2$ between both sets of parameters $\Delta\chi^2\equiv\chi^2(\theta_{\rm fid})-\chi^2(\theta_{\rm bf})$. $\nu_{\rm eff}$ is then given by the median of $\Delta\chi^2$ for several realizations of the synthetic data vector. Through this method we find that $\nu_{\rm eff}$ is well approximated by $\nu_{\rm eff}\simeq N_{\rm data} - (N_b+3)$, where $N_b$ is the number of bias parameters in the model and $N_{\rm data}$ is the number of data points. Each bias parameter is effectively an independent parameter with a broad flat prior affecting the model at the linear level, and thus should add $+1$ to the total $\nu_{\rm eff}$. The contribution from all other parameters is effectively $\Delta\nu_{\rm eff}=3$, either due to their tight priors (in the case of calibrated nuisance parameters), or their limited or correlated impact on the predicted cosmic shear and galaxy clustering observables.

    The main result of this analysis is shown in Fig.~\ref{fig:nested_0}, which presents the constraints on the four cosmological parameters $\Omega_b$, $\Omega_c$, $S_8\equiv\sigma_8(\Omega_m/0.3)^{1/2}$, and $n_s$, for both bias models, with the HEFT results shown for $k_{\rm max}=0.3$ \iMpc (fiducial case) and $k_{\rm max}=0.5$ \iMpc. In the $(\Omega_m,S_8)$ projection, we observe a notable improvement in the constraints on $\Omega_m$. The 68\% confidence interval shrinks from $\Omega_m = 0.298^{+0.033}_{-0.045}$ in the linear bias case to $\Omega_m = 0.273^{+0.030}_{-0.036}$ (15\% improvement from DY1) and $\Omega_m=0.266^{+0.024}_{-0.027}$ (35\% improvement from DY1) using HEFT with $k_{\rm max}=0.3$ \iMpc and $0.5$ \iMpc respectively. The improvement on $S_8$ is less striking (about 10\%). The figure also shows that the scalar spectral index $n_s$ is pushed significantly towards its lower prior bound. As we show in Section \ref{ssec:res.tests}, this is most likely due to the incorrect parameter dependence of the power spectrum templates used, although this does not alter the results found for $(\Omega_m,S_8)$. These results are also summarized in Table~\ref{tab:constraints}, together with the constraints founds for all other data and model configurations explored here.

    \begin{figure}[ht]
      \centering  
      \includegraphics[width=0.85\textwidth]{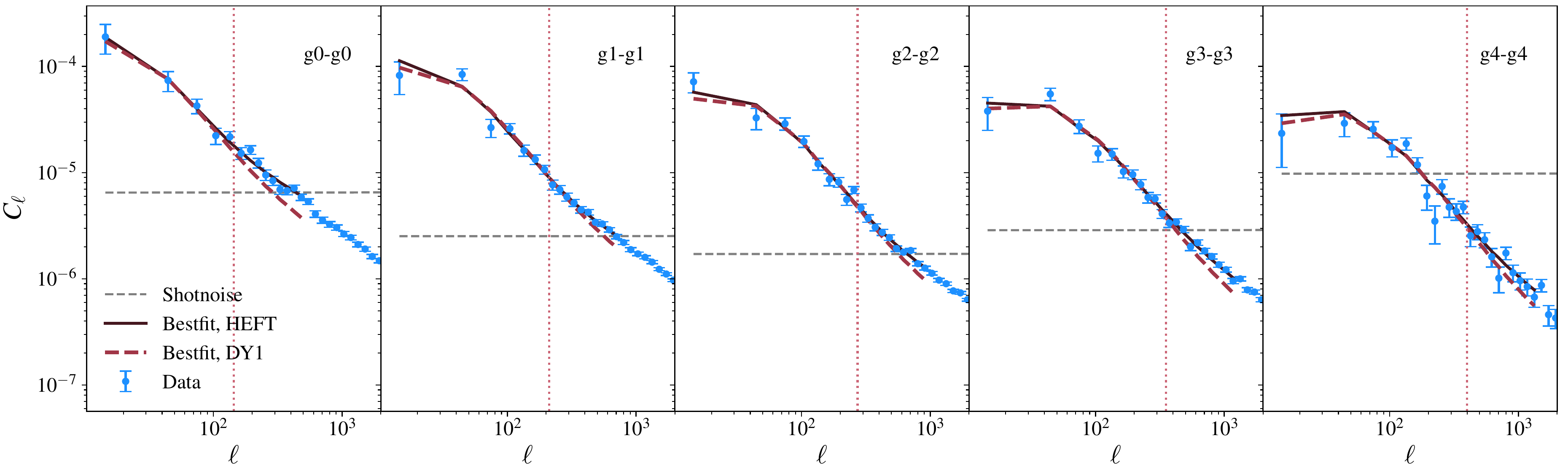} \\
      \includegraphics[width=0.9\textwidth]{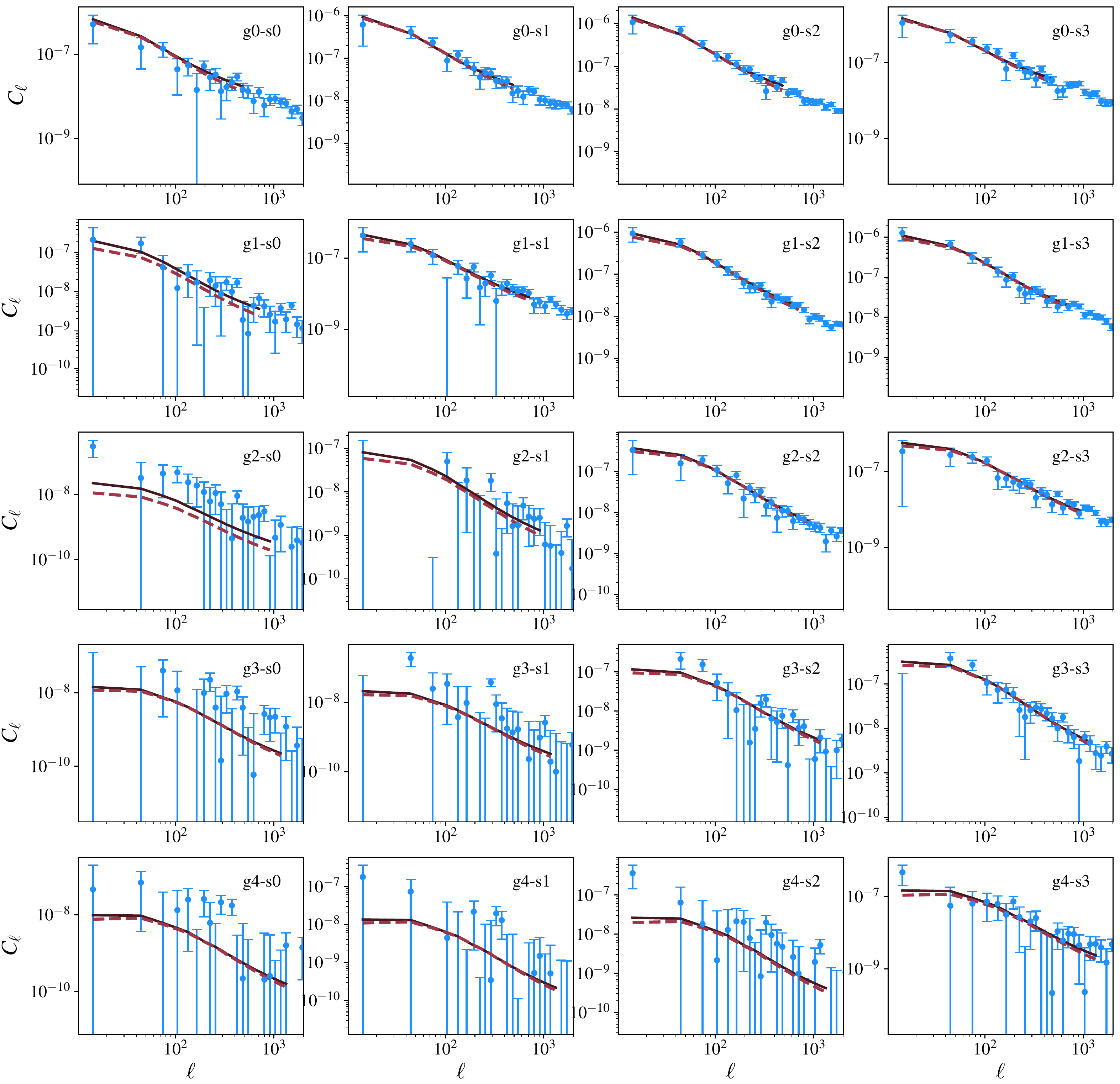}
      \caption{Measured galaxy-galaxy (``g$i$-g$j$'', top panel) and galaxy-shear (``g$i$-s$j$'', bottom panel) power spectra (blue). Here, $i$ and $j$ denote the different tomographic bins (see Fig.~\ref{fig:data}). The solid black and dashed red lines show the theoretical predictions adopting the Hybrid EFT (HEFT) and the DY1 linear bias models respectively. Although we show these predictions up to scales corresponding to $k_{\rm max} = 0.5$ \iMpc, the red dotted vertical line shows the scale cut $k_{\rm max} = 0.15$ \iMpc{} used for the linear bias analysis. The gray dashed horizontal line indicates the Poisson shot-noise of the autocorrelations.}\label{fig:bestfit}
    \end{figure}
    
    It is important to note that, although our results are based on an independent reanalysis of the DY1 dataset in Fourier space, our constraints using a linear bias model are in excellent agreement with those reported by DES \citep{1708.01530}. The estimated power spectra and best-fit predictions for the linear bias and HEFT model are shown in Fig. \ref{fig:bestfit}. Results are shown for the galaxy auto-correlations (labelled $gg$ here) and the cross-correlations with cosmic shear (labelled $gs$). Overall the agreement on large scales is good. The HEFT bias model is able to describe the data on small scales, while the linear model does not capture the small-scale clustering. In the case of the $gs$ we find specific cross-correlations that the model has difficulty fitting, particularly in the lower-left part of the figure, corresponding to cross-correlations where a significant fraction of the lens sample lies behind the source bin. This is likely due to residual systematics in the characterization of the source and lens redshift distributions that are not well captured by the nuisance parameters. Nevertheless, the overall goodness of fit is acceptable, with PTEs above $4.5\%$ (see Section \ref{ssec:res.tests}). The upper panel of Fig. \ref{fig:bestfit} shows, in gray, the shot-noise contribution to the clustering auto-correlations. As we go to higher redshifts, the extended scale range used here lies partially within the noise-dominated regime, limiting the amount of information that can be extracted from the small-scale regime. The use of denser samples, at the cost of broader photomeric redshift uncertainties will likely benefit photometric clustering analyses making use of mildly non-linear scales \citep{2011.03411}.
    \begin{figure}[ht]
      \centering  
      \includegraphics[width=0.90\textwidth]{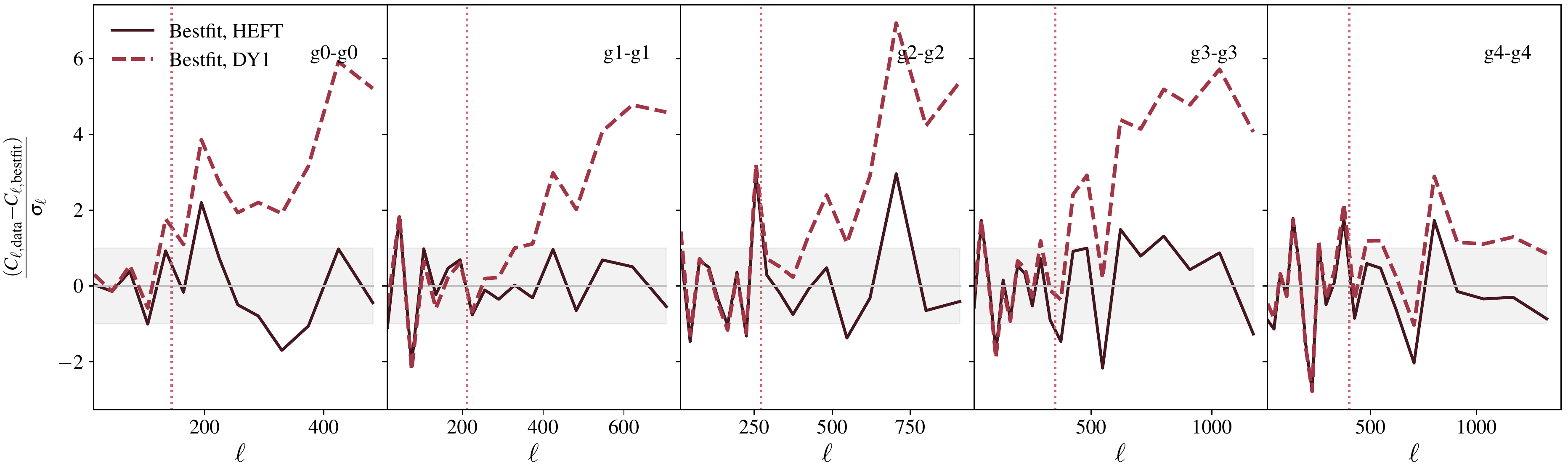} \\
      \includegraphics[width=0.90\textwidth]{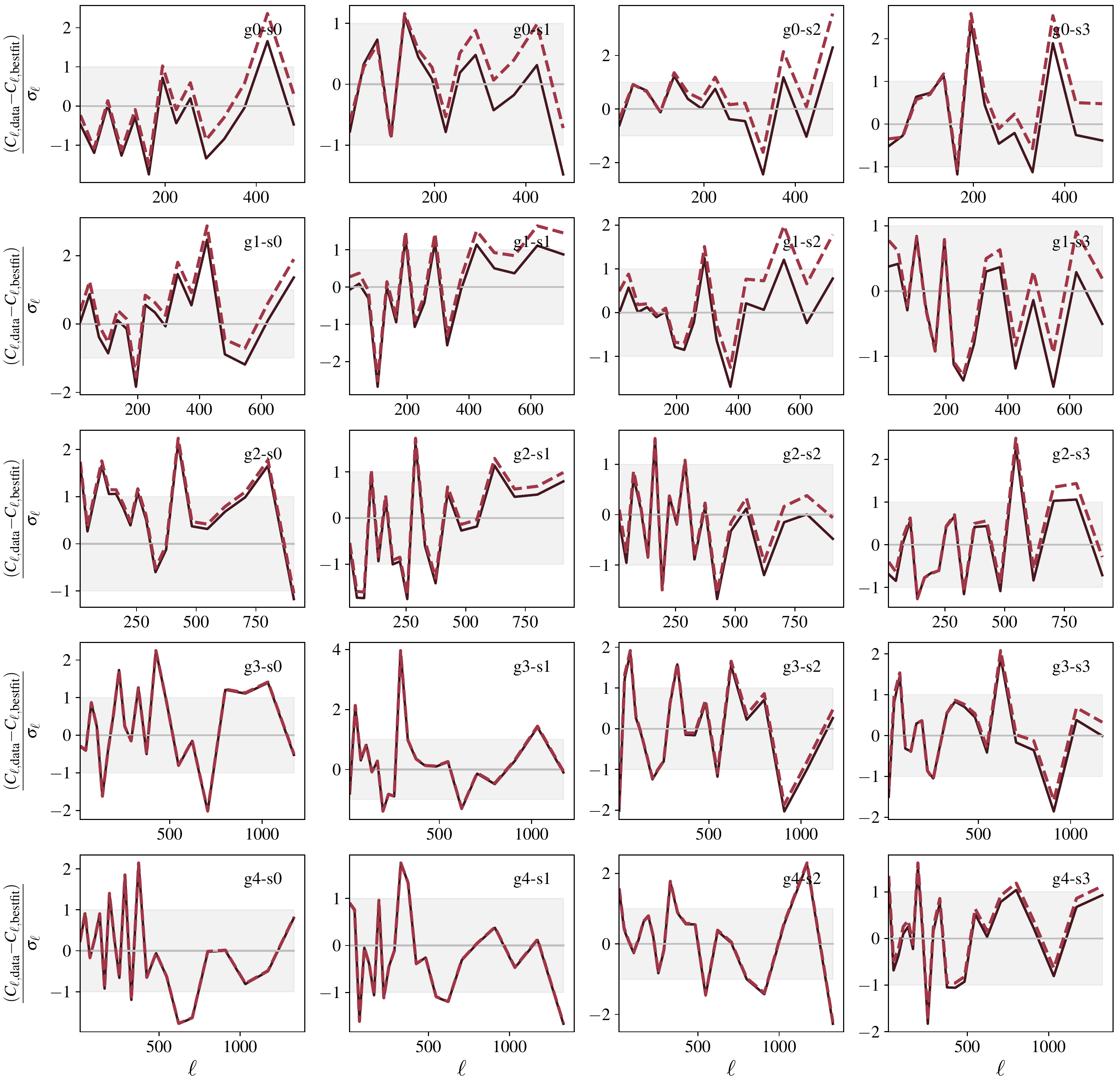}
      \caption{As in Fig. \ref{fig:bestfit}, we show the power spectrum residuals, for the galaxy-galaxy and shear-shear power spectra normalized by their 68\% uncertainties. In solid black and dashed red, we show the residuals for the Hybrid EFT (HEFT) and the DY1 linear bias models respectively. Both agree on large scales, but the linear model is not able to fit the clustering data on scales smaller than the $0.15$ \iMpc{} scale cut (marked by the vertical dotted red line).}\label{fig:chi2_bestfit}
    \end{figure}

    Fig.~\ref{fig:chi2_bestfit} shows the power spectrum residuals as a fraction of the $1\sigma$ uncertainties in the same cases. The linear bias analysis assumes $k_{\rm max} = 0.15$ \iMpc, while the HEFT model has $k_{\rm max} = 0.5$ \iMpc. In the majority of cases, the data lie within 1-2$\sigma$ of the theory in their respective scale ranges, with a small number of exceptions that do not spoil the overall goodness of fit of the HEFT and linear bias models. Importantly, the linear DY1 model shows a significant deviation from the $gg$ data on small scales, which HEFT is able to capture adequately.

    We next explore our results as a function of the scale cut $k_{\rm max}$. The recently developed HEFT $N$-body emulators \citep{2101.11014,2021arXiv210112187Z} was found to fit the halo power spectrum to sub-percent accuracy down to scales $k_{\rm max}\sim0.6$ \hMpc, drilling significantly deeper into the non-linear regime, where various assembly bias effects, related to halo concentration, occupation, local environment, and spin, are known to affect the clustering properties \citep{astro-ph/0108151,astro-ph/0506510,astro-ph/0512416,1911.02610}. In particular, the hybrid approach has been shown to be effective in describing the clustering of more complex tracer populations (see Section 6.3 of \citep{2101.11014}), whereas more traditional methods lead to errors larger than 1\% at $k \approx 0.2$ \iMpc. The improvement in final constraints due to the additional information gained from these modes is offset by the extra freedom allowed by the HEFT model, and eventually limited by the ability of the model to describe the clustering of galaxies. Therefore it is interesting to explore the evolution of the parameter uncertainties and goodness of fit with $k_{\rm max}$.

\begin{figure}[ht]
      \centering  
      \includegraphics[width=.7\textwidth]{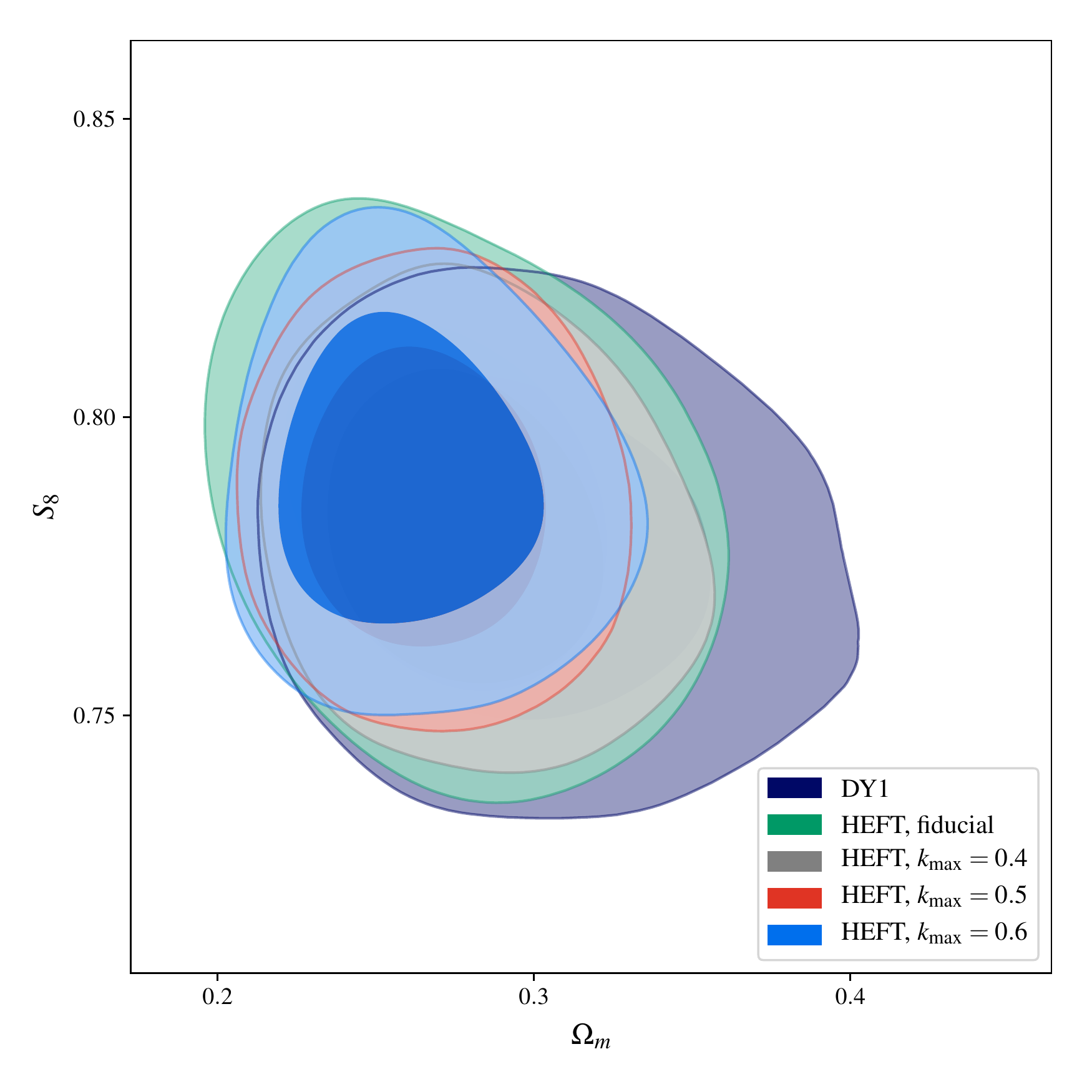}
      \caption{Constraints on the matter fraction $\Omega_m$, and the amplitude of matter fluctuations, parametrized by $S_8\equiv\sigma_8\sqrt{\Omega_m/0.3}$. Results are shown using the linear bias model with $k_{\rm max}=0.15$ \iMpc (dark blue), and the HEFT model with $k_{\rm max} = 0.3, \ 0.4, \ 0.5, \ 0.6$ \iMpc, (green, gray, red and blue respectively). As we go to higher $k_{\rm max}$, the hybrid approach is able to extract additional cosmological information, especially on $\Omega_m$.}\label{fig:money}
    \end{figure}

\begin{figure}[ht]
      \centering  
      \includegraphics[width=0.8\textwidth]{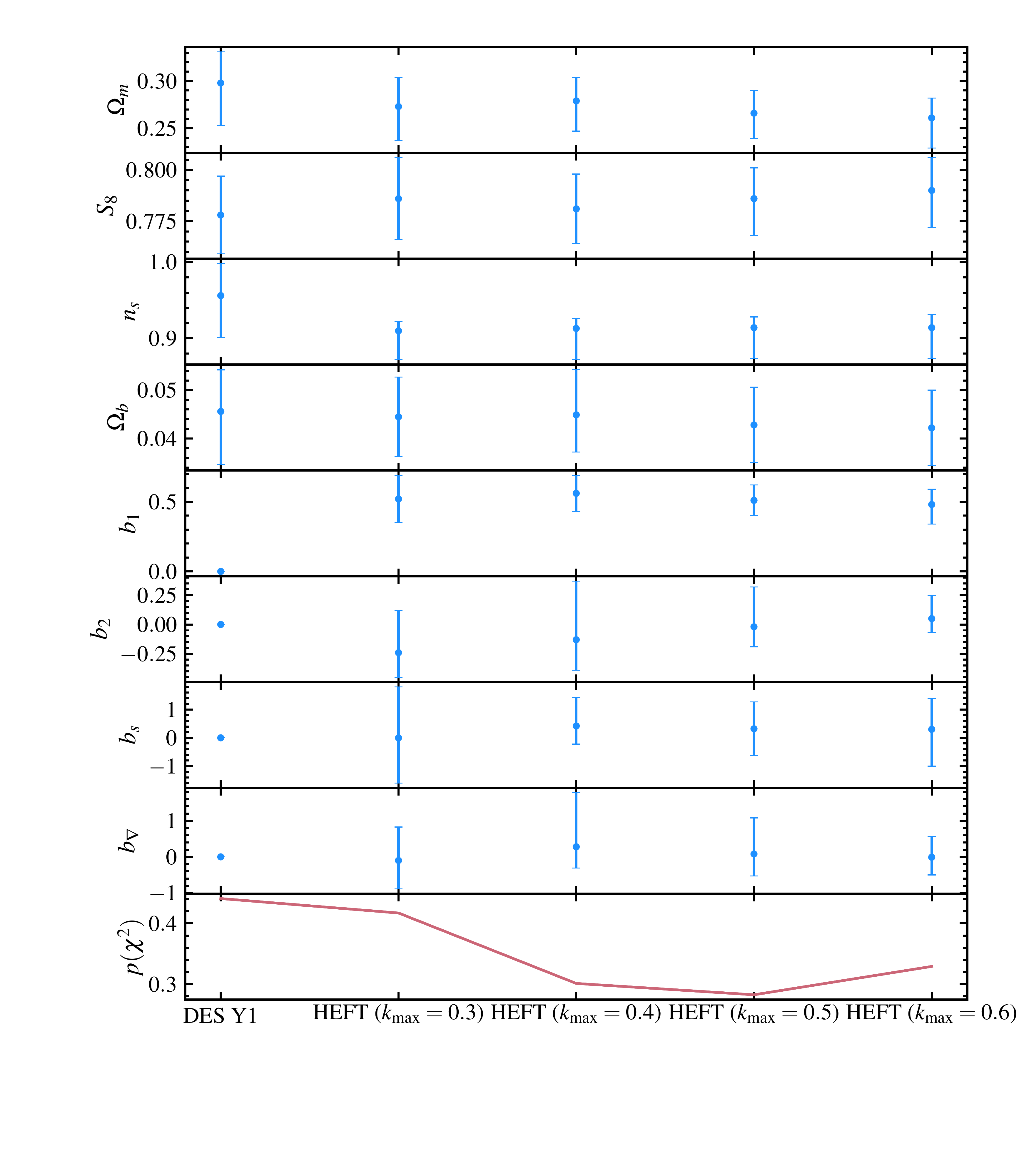}
      \caption{68\% confidence level constraints on four cosmological parameters ($\Omega_m$, $S_8$, $n_s$, $\Omega_b$) and four bias parameters ($b_1$, $b_2$, $b_s$, $b_\nabla$) for the median tomographic redshift bin of the DES Y1 galaxy sources (see Fig.~\ref{fig:data}). The lowermost plot shows the best-fit $\chi^2$ PTEs for the 5 models considered: the linear bias model of the DY1 analysis \cite{1708.01530} and the hybrid model (HEFT) with $k_{\rm max} = 0.3$, 0.4, 0.5, and 0.6 \iMpc. Although the PTEs decrease, as we go to higher $k_{\rm max}$, they remain above $30\%$ throughout. Overall the parameter constraints get tighter, as we go from left to right, implying that the hybrid approach is able to place stricter constraints on both the cosmological parameters (most noticeably $\Omega_m$) and  the bias parameters. Note that the Lagrangian bias parameters are set to zero in the DES Y1 linear model.}\label{fig:evolve}
    \end{figure}

    In Fig.~\ref{fig:money}, we present the joint constraints on $\Omega_m$ and $S_8$ for the HEFT model with $k_{\rm max}$ in the range $[0.3,0.6]$ \iMpc, compared with the constraints found with the linear bias model up to $k_{\rm max}=0.15$ \iMpc. As found before, the uncertainties, particularly on $\Omega_m$, shrink steadily as $k_{\rm max}$ is increased. This improvement, however, seems to asymptote on scales $k_{\rm max}=0.6$ \iMpc, where we recover constraints essentially equivalent to the $k_{\rm max}=0.5$ \iMpc{} case, in spite of adding 35 additional data points. These results are also summarized in Fig. \ref{fig:evolve}, which shows the marginalized constraints on cosmological and bias parameters as a function of bias model and $k_{\rm max}$. The constraints found for different models and scale cuts are broadly consistent with each other. Other than $\Omega_m$ and $S_8$, all other parameters do not benefit significantly from the extended scale range. The bias parameters, however, particularly $b_2$ and $b_\nabla$, are significantly better constrained by the small-scale modes.  
    
    The bottom panel of Figure \ref{fig:evolve} shows the $\chi^2$ PTE for the different cases explored here. Based on the results of \cite{2101.11014} we do not expect the HEFT model to be valid far beyond $k_{\rm max}\simeq0.6$ \hMpc. Nevertheless, we find that the HEFT model is able to describe the data down to the smallest scale explored, which expressed in the \emph{little-h} units of \cite{2101.11014} corresponds to a bone shaking $k_{\rm max}\simeq0.86$ \hMpc. Of course, this statement depends on the statistical power of the data used, and will likely change with future more precise datasets.
    
    \begin{table}
      \small
      \begin{center}
        \begin{tabular}{| l | c | c | c | c | c |}
          \hline\hline
          Model & $\chi^2/\nu_{\rm eff}$ & $S_8$ & $\Omega_m$ & $n_s$ & $H_0$ \\ [0.5ex]
          \hline
DY1 & 470.9/467 & $0.778\pm 0.019$ & $0.298^{+0.033}_{-0.045}$ & $0.956^{+0.042}_{-0.055}$ & --  \\ [1ex]
HEFT, fiducial & 583.5/577 & $0.786\pm 0.020$ & $0.273^{+0.030}_{-0.036}$ & $0.910^{+0.012}_{-0.038}$ & --  \\ [1ex]
HEFT, $k_{\rm max} = 0.4$ & 650.0/632 & $0.781\pm 0.017$ & $0.279^{+0.025}_{-0.032}$ & $0.913^{+0.013}_{-0.041}$ & --  \\ [1ex]
HEFT, $k_{\rm max} = 0.5$ & 702.8/682 & $0.786^{+0.015}_{-0.018}$ & $0.266^{+0.024}_{-0.027}$ & $0.914^{+0.014}_{-0.040}$ & --  \\ [1ex]
HEFT, $k_{\rm max} = 0.6$ & 733.2/717 & $0.790^{+0.016}_{-0.018}$ & $0.261^{+0.021}_{-0.032}$ & $0.914^{+0.017}_{-0.040}$ & --  \\ [1ex]
HEFT, $k_{\rm max} = 0.15$ & 458.8/452 & $0.785\pm 0.020$ & $0.294^{+0.037}_{-0.050}$ & $0.916^{+0.017}_{-0.042}$ & --  \\ [1ex]
HEFT, fixed $P_{ij}(k)$ & 585.4/577 & $0.788\pm 0.021$ & $0.265^{+0.023}_{-0.026}$ & $0.920^{+0.024}_{-0.039}$ & --  \\ [1ex]
HEFT, marg. stochastic & 585.0/577 & $0.784\pm 0.019$ & $0.279^{+0.028}_{-0.039}$ & $0.913^{+0.015}_{-0.039}$ & --  \\ [1ex]
HEFT, {$b_1$, $b_2$, $b_s$} & 584.4/582 & $0.782\pm 0.018$ & $0.270^{+0.026}_{-0.034}$ & $0.925^{+0.021}_{-0.049}$ & --  \\ [1ex]
HEFT, {$b_1$, $b_2$} & 589.2/587 & $0.775\pm 0.017$ & $0.267^{+0.023}_{-0.033}$ & $0.945^{+0.034}_{-0.053}$ & --  \\ [1ex]
DY1, $gg$, $gs$ & 258.3/227 & $0.781\pm 0.042$ & $0.279^{+0.031}_{-0.061}$ & $0.995^{+0.074}_{-0.026}$ & --  \\ [1ex]
HEFT, $gg$, $gs$ & 382.9/337 & $0.777^{+0.032}_{-0.038}$ & $0.299^{+0.038}_{-0.045}$ & $0.9132^{+0.0099}_{-0.043}$ & --  \\ [1ex]
DY1, $H_0$ & 470.3/467 & $0.777\pm 0.019$ & $0.299^{+0.036}_{-0.057}$ & $0.960^{+0.043}_{-0.061}$ & $68.6\pm 6.6$ \\ [1ex]
HEFT, $H_0$, $k_{\rm max} = 0.5$ & 703.6/682 & $0.785\pm 0.017$ & $0.264^{+0.025}_{-0.032}$ & $0.913^{+0.015}_{-0.040}$ & $70.7^{+3.0}_{-3.5}$ \\ [1ex]
          \hline
          \hline
        \end{tabular}
      \end{center}
      \caption{Constraints (68\% confidence level) on the cosmological parameters $S_8$, $\Omega_m$, and $n_s$ for the different data and model configurations considered in this study. We also list the best-fit $\chi^2$ values and effective degrees of freedom $\nu_{\rm eff}$ for each case as a measure of goodness of fit. The definition of $\nu_{\rm eff}$ is discussed in the main text. Rows marked ``DY1'' use a linear bias parametrization, while the others use the HEFT model. The main result is an improvement of $\sim(35\%,10\%)$ in the parameter uncertainties for $(\Omega_m,S_8)$ when using the HEFT model on an extended range of scales. The quoted values of the scale cut $k_{\rm max}$ are in units of \iMpc.} \label{tab:constraints}
    \end{table}

  \subsection{Tests and validation}\label{ssec:res.tests}
    To validate the results presented in the previous section we have carried out a number of additional tests, exploring the contents of the HEFT bias model, the parameter dependence of the power spectrum templates used, and the impact of stochastic bias terms.

    \subsubsection{The HEFT ingredients}\label{sssec:res.tests.bias}
      The HEFT model adds 3 additional free parameters compared with the linear bias parametrization. However, some of these additional degrees of freedom may not be necessary to describe the data on a given range of scales, and it is therefore interesting to explore the possibility of simplifying the model by excluding some of these parameters. In particular, the tidal bias and non-local bias terms ($b_s$ and $b_\nabla$) describe the impact of the local tidal forces, and of physical processes on scales smaller than the characteristic scale for galaxy formation (e.g. the Lagrangian halo size), both of which are arguably subdominant to the impact of the local environmental density \citep{1406.4159,1801.04878}.
      
      To explore this, we repeated our analysis removing $b_\nabla$ alone and together with $b_s$ in our fiducial case with $k_{\rm max}=0.3$ \iMpc. The results from this exercise are summarized in Table~\ref{tab:constraints}. The full-bias model does not provide any significant improvement in the fit ($\Delta\chi^2=(0.9,5.7)$ for 5 and 10 additional parameters respectively). Nevertheless, we find compatible constraints in all cases, which reassures that the additional complexity of the full HEFT model does not degrade the final cosmological constraints significantly. This is to be expected, since non-local effects are likely subdominant these scales.

    \subsubsection{The parameter dependence of HEFT templates}\label{sssec:res.tests.pardep}
      One of the caveats of the analysis carried out here is the simplified method used to characterize the dependence of the HEFT power spectrum templates on cosmological parameters. As described in Section \ref{ssec:theory.abacus}, the ratios of the templates with respect to the matter power spectrum are only allowed to vary linearly with respect to the cosmological parameters around the fiducial cosmology of the \abacus{} suite.

      To quantify the impact of this approximation we repeat our analysis for $k_{\rm max}=0.3$ \iMpc{} removing this linear parameter dependence altogether, and assuming constant power spectrum template ratios. In this cruder approximation, all parameter dependence of the power spectrum templates is captured by the matter power spectrum. The results, listed in Table \ref{tab:constraints}, show that the constraints on the primary large-scale structure parameters $S_8$ and $\Omega_m$ are in good agreement with those found in the fiducial analysis, and that the parameter dependence of the template ratios does not improve them significantly. The reduction in uncertainties found before is therefore not artificially caused by incorrectly accounting for this parameter dependence.

      However, we find that the constraints on $n_s$ in both cases, although still compatible at $1\sigma$, show significantly different uncertainties. We believe this effect is indeed caused by an inaccurate modeling of the $n_s$ dependence, where the linear approximation assumed here breaks down more readily. To confirm that the improved uncertainties on $n_s$ are indeed artificial, and not caused by the additional small-scale information enabled by the HEFT model, we repeat our analysis in the reduced scale range $k<0.15$ \iMpc\ used with the linear bias model. The results of this test are also listed in Table \ref{tab:constraints}, and show that, even in this reduced range, we obtain tighter constraints on $n_s$ than those found with the simpler linear bias model. This confirms that the dependence of the power spectrum template ratios on $n_s$ is not correctly captured with our simplified setup. All future analyses using the HEFT model should therefore rely on full emulators where this dependence is correctly captured. We emphasize, however, that the constraints found in the $(\Omega_m,S_8)$ plane are unaffected by this approximation, and indeed the HEFT results in the reduced scale range agree rather well with those found for the linear bias model for these parameters.

      Another aspect of the cosmological dependence not captured by our current implementation is the dependence on the local expansion rate $H_0$. Since $\theta_\ast$ is the same for all the \abacus{} simulations used here, $H_0$ was treated as a derived parameter, determined in terms of the other cosmological parameters by fixing $\theta_\ast$ to the \abacus{} value. Although we do not expect this to impact our results significantly, since the DY1 analysis was not able to constrain $H_0$, we have repeated our analysis treating $H_0$ as a free parameter. This was done for the DY1 linear bias model with $k_{\rm max}=0.15$ \iMpc, and for the HEFT model with $k_{\rm max}=0.5$  \iMpc. Due to the limitation we just described, we are not able to account for this parameter dependence in the HEFT power spectrum ratios, and therefore the dependence is only included in the \texttt{HALOFIT} matter power spectrum. The results, listed in Table \ref{tab:constraints}, show that the constraints on $(\Omega_m,S_8)$ are not affected by the additional parameter freedom. Interestingly, we find that we are able to recover tighter constraints on $H_0$ ($\sigma(H_0)\simeq3.3\,{\rm km}\,s^{-1}\,{\rm Mpc}^{-1}$) than those found in the linear analysis ($\sigma(H_0)\simeq6.6\,{\rm km}\,s^{-1}\,{\rm Mpc}^{-1}$). Since the dependence on $H_0$ may be imperfectly captured by our implementation, we cannot make a claim regarding this result. However, the possibility of constraining $H_0$ from the joint analysis of galaxy clustering and weak lensing on the smaller scales enabled by the HEFT bias model, should be explored in the future.

    \subsubsection{Stochasticity}\label{sssec:res.tests.stochastic}
      As described in Section \ref{ssec:theory.heft}, we expect the galaxy overdensity to receive a stochastic contribution sourced by the small-scale physics governing galaxy formation and evolution. This contribution should dominate the galaxy power spectrum on small scales and, as shown in Fig. \ref{fig:data}, dominates the deterministic signal on angular scales $\ell\gtrsim1000$. By inspecting the measured power spectra on these scales, we find that the stochastic contribution is well described by a Poisson-like process, with $C^{\rm SN}_\ell=1/\bar{n}_\Omega$ after accounting for the effects of inhomogeneous sky coverage, where $\bar{n}_\Omega$ is the number of galaxies per steradian. Nevertheless, to allow for possible residual stochastic contributions, we repeat our analysis allowing for an additional free amplitude parameter $A_\epsilon$ per redshift bin multiplying this shot-noise contribution, and marginalize over it with a conservative 10\% Gaussian prior centered on $A_\epsilon=1$. Since $A_\epsilon$ is a linear parameter, this marginalization is done analytically by simply modifying the covariance matrix as:
      \begin{equation}
        {\rm Cov}_{\ell\ell'}\longrightarrow
        {\rm Cov}_{\ell\ell'}+\sigma_A^2C_\ell^{\rm SN}C_{\ell'}^{\rm SN},
      \end{equation}
      where $\sigma_A=0.1$.

      The result of this test, listed in Table \ref{tab:constraints}, shows that the effect of residual stochastic contributions is negligible, both in the best-fit parameters and their uncertainty.

    \subsubsection{Excluding shear}\label{sssec:res.tests.ss}
      The three main components of the data vector used here are the galaxy-galaxy, galaxy-shear, and shear-shear correlations ($gg$, $gs$ and $ss$ respectively). Of these, $ss$ is not sensitive to any modeling improvement brought about by the use of the HEFT model. In order to better isolate the parameter constraint enhancement, and the goodness of fit associated with the HEFT model, we repeated our analysis excluding the $ss$ component from the data. This was done both for the DY1 linear bias model with $k<0.15$ \iMpc, and for the HEFT model with $k_{\rm max}=0.3$ \iMpc. The results are shown in Table \ref{tab:constraints}.

      When excluding the shear-shear data, the uncertainty on $S_8$ grows by $\sim60\%$ for both bias models, while the constraints on $\Omega_m$ grow by about $\sim20\%$. This shows that the inclusion of galaxy clustering information plays a vital role in constraining $\Omega_m$ \citep{1708.01530}. The relative improvement in constraints associated with the HEFT model is similar in both cases. Excluding the effects of intrinsic alignments, cosmic shear is an unbiased tracer of the matter fluctuations. Thus, in spite of its significantly lower signal-to-noise ratio compared with galaxy clustering, the shear power spectrum is able to obtain comparatively tighter constraints.

      It is worth noting that the goodness of fit of the $gg$-$gs$ sector is notably lower for both bias models than that of the full data vector, with a minimum PTE of 4.5\%. Although the statistical significance is not high, this may be a sign of unmodelled systematics in the data. A visual inspection of the data suggests that it is impossible for the model to fit some of the galaxy-shear power spectra corresponding to lens samples partially behind the sources (e.g. ``g2-s0'' or ``g3-s0'' in Fig. \ref{fig:data}). This is a feature of both the linear and HEFT models, and could be ascribed to an imperfect model of the source or lens redshift distributions.

\section{Discussion \& Conclusions}\label{sec:conclusions}
  This paper presents the first application of the Hybrid EFT (HEFT) bias model of \cite{1910.07097} to observational data. In particular, we re-analyzed the combination of lensing shear and projected galaxy clustering from the DES Y1 data release. This methods holds promise to be the method of choice for next generation surveys, since it efficiently combines the reach of $N$-body simulations deep into the non-linear regime with the analytical exactness and theoretical control of bias expansions. We note that photometric surveys are especially well suited for the first  application of this method, since we do not require a precise model for redshift-space distortions. 

  We calibrated our HEFT basis power spectra from the \abacus{} $N$-body simulations, smoothly interpolating into large-scale analytical solutions and using a linear expansion around the fiducial model to account for the cosmology dependence. In this preliminary work we have focused on the goodness-of-fit as basic quantity by which to choose whether a model produces a reliable fit.

  Our results can be summarized as follows:
  \begin{itemize}
    \item HEFT models offer a good fit to the data. Nominally, the $\chi^2$ is good all the way to $k_{\rm max} = 0.6$ \iMpc
    \item HEFT models offer a stable fit to the data. While the uncertainties decrease, the inferred values of all parameters remain consistent as we push to higher $k_{\rm max}$;
    \item Given the enormous increase in signal-to-noise, as we increase the $k_{\rm max}$, the improvement on the cosmological parameters of interest is rather modest. The standard interpretation is that all the new information is going mostly into determining the bias parameters rather than the cosmology. In some sense this method offers ``graceful transition to ignorance'', where the information in the 2-point function is being exhausted and the results do not depend strongly on the choice of scale cut. This observation also implies that a 3-point function over the same scales could break further degeneracies and bring concrete improvements in our determination of cosmological parameters.
    \item While the constraint on $S_8$ improves modestly (about $\sim 10$\%) for $k_{\rm max}=0.5\,$\iMpc,  we do notice a significant improvement in the uncertainty on $\Omega_m$, which shrinks by about 35\% to $\Omega_m=0.266^{+0.024}_{-0.027}$. This implies that HEFT starts to break the degeneracy in the less constraining direction of the ``weak-lensing banana'' contour on the $\sigma_8-\Omega_m$-plane. Interestingly, our results are $\sim2\sigma$ away from the central value found by \planck, $\Omega_m=0.315\pm0.007$ \citep{1807.06209}.
    \item In our fits, we have left $\Omega_b$, $n_s$ and $H_0$ (with fixed angular diameter distance to the last scattering)  as free parameters. While it is not expected that the combination of cosmic shear and galaxy clustering at the level of two-point functions can measure these with significant precision, it is nevertheless re-assuring that the parameters were bound by the data, rather than the prior and that Planck-determined values lie well withing the posteriors spanned by them.
    \item While analyzing the data, we noticed that the $\chi^2$ values for the $gg$ and $gs$ power spectra are somewhat high. However, the statistical significance is small when accounting for the effective number of fitted degrees of freedom.
  \end{itemize}

  We reiterate that the work presented here is fundamentally exploratory in nature and our constraints should not be taken to be as robust as those supported by extensive testing on realistic mock datasets.  There are numerous steps where our method can be improved.

  Most importantly, the exact $k_{\rm max}$ limits should be obtained based on mock galaxy catalogs populated with realistic distribution of galaxies, rather than relying solely on the goodness of fit. Ideally, such mock catalogs should be based on an independent suite of $N$-body simulations spanning plausible cosmologies and recipes for populating halos with galaxies. Methodologically, the way templates are created leaves several open questions in terms of how basis spectra are generated and interpolated. At the moment we have chosen the pivot points for interpolation between analytic and $N$-body solutions essentially ``by hand''. Basis spectra involving $\delta^2_L$ and $\nabla\delta_L$ are noisy and may show a weak dependence on the smoothing scale (not used in this paper). The sample variance in the basis spectra is poorly understood. As we were building our prediction scheme, independent emulators were developed by \cite{2101.11014} and \cite{2021arXiv210112187Z}. Repeating our analysis by replacing our simple linear parametrization (see Eq. \ref{eq:pktaylor} with such an emulator would be a very interesting cross-check that would quantify the systematic errors in our theory model. We leave this, and the other caveats listed in this paragraph, for future work. As the power of the HEFT approach becomes apparent, we expect that the coming years will bring about more sophisticated methods to measure and interpolate the relevant quantities from $N$-body simulations.

  Despite these caveats, HEFT worked on DES Y1 data ``out of the box'', producing a good and stable fit without any tweaking. Our results confirm the robustness of DES Y1 data and demonstrate the promise of the upcoming generation of photometric galaxy surveys when analyzed with state-of-the-art theory prediction tools.

\section*{Acknowledgements}
We would like to thank Shi-Fan Chen, Lehman Garrison, Daniel Eisenstein, Yu Feng, Heather Kelly, Chirag Modi, and Martin White for fruitful conversations and helpful advice, which was much needed at the initial stages of the project.

CGG acknowledge support from the European Research Council Grant No:  693024 and the Beecroft Trust. DA is supported by the Science and Technology Facilities Council through an Ernest Rutherford Fellowship, grant reference ST/P004474.  This research used resources of the National Energy Research Scientific Computing Center (NERSC), a U.S. Department of Energy Office of Science User Facility located at Lawrence Berkeley National Laboratory, for computing the power spectrum templates, $P_{\alpha \beta}(k)$, which constitute a sizeable portion of the project. We also made extensive use of computational resources at the University of Oxford Department of Physics, funded by the John Fell Oxford University Press Research Fund.

We made extensive use of the {\tt numpy} \citep{oliphant2006guide, van2011numpy}, {\tt scipy} \citep{2020SciPy-NMeth}, {\tt astropy} \citep{astropy:2013, astropy:2018}, {\tt healpy} \citep{Zonca2019}, \texttt{nbodykit} \citep{2018AJ....156..160H}, \texttt{GetDist} \citep{2019arXiv191013970L}, \texttt{pyccl} \citep{1812.05995}, \texttt{MontePython} \citep{2019PDU....24..260B}, \texttt{velocileptors} \citep{2005.00523}, and {\tt matplotlib} \citep{Hunter:2007} python packages.

This paper makes use of software developed for the Large Synoptic Survey Telescope. We thank the LSST Project for making their code available as free software at \url{http://dm.lsst.org}.

This project used public archival data from the Dark Energy Survey (DES). Funding for the DES Projects has been provided by the U.S. Department of Energy, the U.S. National Science Foundation, the Ministry of Science and Education of Spain, the Science and Technology FacilitiesCouncil of the United Kingdom, the Higher Education Funding Council for England, the National Center for Supercomputing Applications at the University of Illinois at Urbana-Champaign, the Kavli Institute of Cosmological Physics at the University of Chicago, the Center for Cosmology and Astro-Particle Physics at the Ohio State University, the Mitchell Institute for Fundamental Physics and Astronomy at Texas A\&M University, Financiadora de Estudos e Projetos, Funda{\c c}{\~a}o Carlos Chagas Filho de Amparo {\`a} Pesquisa do Estado do Rio de Janeiro, Conselho Nacional de Desenvolvimento Cient{\'i}fico e Tecnol{\'o}gico and the Minist{\'e}rio da Ci{\^e}ncia, Tecnologia e Inova{\c c}{\~a}o, the Deutsche Forschungsgemeinschaft, and the Collaborating Institutions in the Dark Energy Survey.
    
The Collaborating Institutions are Argonne National Laboratory, the University of California at Santa Cruz, the University of Cambridge, Centro de Investigaciones Energ{\'e}ticas, Medioambientales y Tecnol{\'o}gicas-Madrid, the University of Chicago, University College London, the DES-Brazil Consortium, the University of Edinburgh, the Eidgen{\"o}ssische Technische Hochschule (ETH) Z{\"u}rich,  Fermi National Accelerator Laboratory, the University of Illinois at Urbana-Champaign, the Institut de Ci{\`e}ncies de l'Espai (IEEC/CSIC), the Institut de F{\'i}sica d'Altes Energies, Lawrence Berkeley National Laboratory, the Ludwig-Maximilians Universit{\"a}t M{\"u}nchen and the associated Excellence Cluster Universe, the University of Michigan, the National Optical Astronomy Observatory, the University of Nottingham, The Ohio State University, the OzDES Membership Consortium, the University of Pennsylvania, the University of Portsmouth, SLAC National Accelerator Laboratory, Stanford University, the University of Sussex, and Texas A\&M University.

Based in part on observations at Cerro Tololo Inter-American Observatory, National Optical Astronomy Observatory, which is operated by the Association of Universities for Research in Astronomy (AURA) under a cooperative agreement with the National Science Foundation.

\bibliography{bibliography,non_ads}

\providecommand{\href}[2]{#2}\begingroup\raggedright\begin{thebibliography}{10}

\bibitem{1910.07097}
C.~{Modi}, S.-F. {Chen} and M.~{White}, \emph{{Simulations and symmetries}},
  \href{https://doi.org/10.1093/mnras/staa251}{\emph{\mnras} {\bfseries 492}
  (2020) 5754} [\href{https://arxiv.org/abs/1910.07097}{{\ttfamily
  1910.07097}}].

\bibitem{1708.01530}
T.~M.~C. {Abbott}, F.~B. {Abdalla}, A.~{Alarcon}, J.~{Aleksi{\'c}}, S.~{Allam},
  S.~{Allen} et~al., \emph{{Dark Energy Survey year 1 results: Cosmological
  constraints from galaxy clustering and weak lensing}},
  \href{https://doi.org/10.1103/PhysRevD.98.043526}{\emph{\prd} {\bfseries 98}
  (2018) 043526} [\href{https://arxiv.org/abs/1708.01530}{{\ttfamily
  1708.01530}}].

\bibitem{1809.09148}
C.~{Hikage}, M.~{Oguri}, T.~{Hamana}, S.~{More}, R.~{Mandelbaum}, M.~{Takada}
  et~al., \emph{{Cosmology from cosmic shear power spectra with Subaru Hyper
  Suprime-Cam first-year data}},
  \href{https://doi.org/10.1093/pasj/psz010}{\emph{\pasj} {\bfseries 71} (2019)
  43} [\href{https://arxiv.org/abs/1809.09148}{{\ttfamily 1809.09148}}].

\bibitem{2007.15632}
C.~{Heymans}, T.~{Tr{\"o}ster}, M.~{Asgari}, C.~{Blake}, H.~{Hildebrandt},
  B.~{Joachimi} et~al., \emph{{KiDS-1000 Cosmology: Multi-probe weak
  gravitational lensing and spectroscopic galaxy clustering constraints}},
  {\emph{arXiv e-prints} (2020) arXiv:2007.15632}
  [\href{https://arxiv.org/abs/2007.15632}{{\ttfamily 2007.15632}}].

\bibitem{1708.01536}
J.~{Elvin-Poole}, M.~{Crocce}, A.~J. {Ross}, T.~{Giannantonio}, E.~{Rozo},
  E.~S. {Rykoff} et~al., \emph{{Dark Energy Survey year 1 results: Galaxy
  clustering for combined probes}},
  \href{https://doi.org/10.1103/PhysRevD.98.042006}{\emph{\prd} {\bfseries 98}
  (2018) 042006} [\href{https://arxiv.org/abs/1708.01536}{{\ttfamily
  1708.01536}}].

\bibitem{1104.1174}
M.~P. {van Daalen}, J.~{Schaye}, C.~M. {Booth} and C.~{Dalla Vecchia},
  \emph{{The effects of galaxy formation on the matter power spectrum: a
  challenge for precision cosmology}},
  \href{https://doi.org/10.1111/j.1365-2966.2011.18981.x}{\emph{\mnras}
  {\bfseries 415} (2011) 3649}
  [\href{https://arxiv.org/abs/1104.1174}{{\ttfamily 1104.1174}}].

\bibitem{1105.1075}
E.~{Semboloni}, H.~{Hoekstra}, J.~{Schaye}, M.~P. {van Daalen} and I.~G.
  {McCarthy}, \emph{{Quantifying the effect of baryon physics on weak lensing
  tomography}},
  \href{https://doi.org/10.1111/j.1365-2966.2011.19385.x}{\emph{\mnras}
  {\bfseries 417} (2011) 2020}
  [\href{https://arxiv.org/abs/1105.1075}{{\ttfamily 1105.1075}}].

\bibitem{1405.7423}
T.~{Eifler}, E.~{Krause}, S.~{Dodelson}, A.~R. {Zentner}, A.~P. {Hearin} and
  N.~Y. {Gnedin}, \emph{{Accounting for baryonic effects in cosmic shear
  tomography: determining a minimal set of nuisance parameters using PCA}},
  \href{https://doi.org/10.1093/mnras/stv2000}{\emph{\mnras} {\bfseries 454}
  (2015) 2451} [\href{https://arxiv.org/abs/1405.7423}{{\ttfamily 1405.7423}}].

\bibitem{1809.01146}
H.-J. {Huang}, T.~{Eifler}, R.~{Mandelbaum} and S.~{Dodelson}, \emph{{Modelling
  baryonic physics in future weak lensing surveys}},
  \href{https://doi.org/10.1093/mnras/stz1714}{\emph{\mnras} {\bfseries 488}
  (2019) 1652} [\href{https://arxiv.org/abs/1809.01146}{{\ttfamily
  1809.01146}}].

\bibitem{1810.08629}
A.~{Schneider}, R.~{Teyssier}, J.~{Stadel}, N.~E. {Chisari}, A.~M.~C. {Le
  Brun}, A.~{Amara} et~al., \emph{{Quantifying baryon effects on the matter
  power spectrum and the weak lensing shear correlation}},
  \href{https://doi.org/10.1088/1475-7516/2019/03/020}{\emph{\jcap} {\bfseries
  2019} (2019) 020} [\href{https://arxiv.org/abs/1810.08629}{{\ttfamily
  1810.08629}}].

\bibitem{1801.08559}
N.~E. {Chisari}, M.~L.~A. {Richardson}, J.~{Devriendt}, Y.~{Dubois},
  A.~{Schneider}, A.~M.~C. {Le Brun} et~al., \emph{{The impact of baryons on
  the matter power spectrum from the Horizon-AGN cosmological hydrodynamical
  simulation}}, \href{https://doi.org/10.1093/mnras/sty2093}{\emph{\mnras}
  {\bfseries 480} (2018) 3962}
  [\href{https://arxiv.org/abs/1801.08559}{{\ttfamily 1801.08559}}].

\bibitem{astro-ph/0206508}
A.~A. {Berlind} and D.~H. {Weinberg}, \emph{{The Halo Occupation Distribution:
  Toward an Empirical Determination of the Relation between Galaxies and
  Mass}}, \href{https://doi.org/10.1086/341469}{\emph{\apj} {\bfseries 575}
  (2002) 587} [\href{https://arxiv.org/abs/astro-ph/0109001}{{\ttfamily
  astro-ph/0109001}}].

\bibitem{astro-ph/0005010}
J.~A. {Peacock} and R.~E. {Smith}, \emph{{Halo occupation numbers and galaxy
  bias}}, \href{https://doi.org/10.1046/j.1365-8711.2000.03779.x}{\emph{\mnras}
  {\bfseries 318} (2000) 1144}
  [\href{https://arxiv.org/abs/astro-ph/0005010}{{\ttfamily
  astro-ph/0005010}}].

\bibitem{1406.7843}
L.~{Senatore}, \emph{{Bias in the effective field theory of large scale
  structures}},
  \href{https://doi.org/10.1088/1475-7516/2015/11/007}{\emph{\jcap} {\bfseries
  2015} (2015) 007} [\href{https://arxiv.org/abs/1406.7843}{{\ttfamily
  1406.7843}}].

\bibitem{1611.09787}
V.~{Desjacques}, D.~{Jeong} and F.~{Schmidt}, \emph{{Large-scale galaxy bias}},
  \href{https://doi.org/10.1016/j.physrep.2017.12.002}{\emph{\physrep}
  {\bfseries 733} (2018) 1} [\href{https://arxiv.org/abs/1611.09787}{{\ttfamily
  1611.09787}}].

\bibitem{1912.08209}
A.~{Nicola}, D.~{Alonso}, J.~{S{\'a}nchez}, A.~{Slosar}, H.~{Awan},
  A.~{Broussard} et~al., \emph{{Tomographic galaxy clustering with the Subaru
  Hyper Suprime-Cam first year public data release}},
  \href{https://doi.org/10.1088/1475-7516/2020/03/044}{\emph{\jcap} {\bfseries
  2020} (2020) 044} [\href{https://arxiv.org/abs/1912.08209}{{\ttfamily
  1912.08209}}].

\bibitem{2010.09717}
A.~{Nicola}, C.~{Garc{\'\i}a-Garc{\'\i}a}, D.~{Alonso}, J.~{Dunkley}, P.~G.
  {Ferreira}, A.~{Slosar} et~al., \emph{{Cosmic shear power spectra in
  practice}}, {\emph{arXiv e-prints} (2020) arXiv:2010.09717}
  [\href{https://arxiv.org/abs/2010.09717}{{\ttfamily 2010.09717}}].

\bibitem{2018ApJS..239...18A}
T.~M.~C. {Abbott}, F.~B. {Abdalla}, S.~{Allam}, A.~{Amara}, J.~{Annis},
  J.~{Asorey} et~al., \emph{{The Dark Energy Survey: Data Release 1}},
  \href{https://doi.org/10.3847/1538-4365/aae9f0}{\emph{\apjs} {\bfseries 239}
  (2018) 18} [\href{https://arxiv.org/abs/1801.03181}{{\ttfamily 1801.03181}}].

\bibitem{1702.02600}
E.~{Huff} and R.~{Mandelbaum}, \emph{{Metacalibration: Direct Self-Calibration
  of Biases in Shear Measurement}}, {\emph{arXiv e-prints} (2017)
  arXiv:1702.02600} [\href{https://arxiv.org/abs/1702.02600}{{\ttfamily
  1702.02600}}].

\bibitem{1702.02601}
E.~S. {Sheldon} and E.~M. {Huff}, \emph{{Practical Weak-lensing Shear
  Measurement with Metacalibration}},
  \href{https://doi.org/10.3847/1538-4357/aa704b}{\emph{\apj} {\bfseries 841}
  (2017) 24} [\href{https://arxiv.org/abs/1702.02601}{{\ttfamily 1702.02601}}].

\bibitem{1708.01533}
J.~{Zuntz}, E.~{Sheldon}, S.~{Samuroff}, M.~A. {Troxel}, M.~{Jarvis},
  N.~{MacCrann} et~al., \emph{{Dark Energy Survey Year 1 results: weak lensing
  shape catalogues}},
  \href{https://doi.org/10.1093/mnras/sty2219}{\emph{\mnras} {\bfseries 481}
  (2018) 1149} [\href{https://arxiv.org/abs/1708.01533}{{\ttfamily
  1708.01533}}].

\bibitem{1708.01532}
B.~{Hoyle}, D.~{Gruen}, G.~M. {Bernstein}, M.~M. {Rau}, J.~{De Vicente}, W.~G.
  {Hartley} et~al., \emph{{Dark Energy Survey Year 1 Results: redshift
  distributions of the weak-lensing source galaxies}},
  \href{https://doi.org/10.1093/mnras/sty957}{\emph{\mnras} {\bfseries 478}
  (2018) 592} [\href{https://arxiv.org/abs/1708.01532}{{\ttfamily
  1708.01532}}].

\bibitem{1809.09603}
D.~{Alonso}, J.~{Sanchez}, A.~{Slosar} and {LSST Dark Energy Science
  Collaboration}, \emph{{A unified pseudo-C$_{{\ensuremath{\ell}}}$
  framework}}, \href{https://doi.org/10.1093/mnras/stz093}{\emph{\mnras}
  {\bfseries 484} (2019) 4127}
  [\href{https://arxiv.org/abs/1809.09603}{{\ttfamily 1809.09603}}].

\bibitem{1906.11765}
C.~{Garc{\'\i}a-Garc{\'\i}a}, D.~{Alonso} and E.~{Bellini}, \emph{{Disconnected
  pseudo-C$_{l}$ covariances for projected large-scale structure data}},
  \href{https://doi.org/10.1088/1475-7516/2019/11/043}{\emph{\jcap} {\bfseries
  2019} (2019) 043} [\href{https://arxiv.org/abs/1906.11765}{{\ttfamily
  1906.11765}}].

\bibitem{2011.06469}
C.~{Doux}, C.~{Chang}, B.~{Jain}, J.~{Blazek}, H.~{Camacho}, X.~{Fang} et~al.,
  \emph{{Consistency of cosmic shear analyses in harmonic and real space}},
  \href{https://doi.org/10.1093/mnras/stab661}{\emph{\mnras} (2021) }
  [\href{https://arxiv.org/abs/2011.06469}{{\ttfamily 2011.06469}}].

\bibitem{astro-ph/0105302}
E.~{Hivon}, K.~M. {G{\'o}rski}, C.~B. {Netterfield}, B.~P. {Crill}, S.~{Prunet}
  and F.~{Hansen}, \emph{{MASTER of the Cosmic Microwave Background Anisotropy
  Power Spectrum: A Fast Method for Statistical Analysis of Large and Complex
  Cosmic Microwave Background Data Sets}},
  \href{https://doi.org/10.1086/338126}{\emph{\apj} {\bfseries 567} (2002) 2}
  [\href{https://arxiv.org/abs/astro-ph/0105302}{{\ttfamily
  astro-ph/0105302}}].

\bibitem{astro-ph/9611174}
M.~{Tegmark}, \emph{{How to measure CMB power spectra without losing
  information}}, \href{https://doi.org/10.1103/PhysRevD.55.5895}{\emph{\prd}
  {\bfseries 55} (1997) 5895}
  [\href{https://arxiv.org/abs/astro-ph/9611174}{{\ttfamily
  astro-ph/9611174}}].

\bibitem{astro-ph/0409513}
K.~M. {G{\'o}rski}, E.~{Hivon}, A.~J. {Banday}, B.~D. {Wand elt}, F.~K.
  {Hansen}, M.~{Reinecke} et~al., \emph{{HEALPix: A Framework for
  High-Resolution Discretization and Fast Analysis of Data Distributed on the
  Sphere}}, \href{https://doi.org/10.1086/427976}{\emph{\apj} {\bfseries 622}
  (2005) 759} [\href{https://arxiv.org/abs/astro-ph/0409513}{{\ttfamily
  astro-ph/0409513}}].

\bibitem{2017MNRAS.467.2085G}
J.~N. {Grieb}, A.~G. {S{\'a}nchez}, S.~{Salazar-Albornoz}, R.~{Scoccimarro},
  M.~{Crocce}, C.~{Dalla Vecchia} et~al., \emph{{The clustering of galaxies in
  the completed SDSS-III Baryon Oscillation Spectroscopic Survey: Cosmological
  implications of the Fourier space wedges of the final sample}},
  \href{https://doi.org/10.1093/mnras/stw3384}{\emph{\mnras} {\bfseries 467}
  (2017) 2085} [\href{https://arxiv.org/abs/1607.03143}{{\ttfamily
  1607.03143}}].

\bibitem{0810.4170}
M.~{Takada} and B.~{Jain}, \emph{{The impact of non-Gaussian errors on weak
  lensing surveys}},
  \href{https://doi.org/10.1111/j.1365-2966.2009.14504.x}{\emph{\mnras}
  {\bfseries 395} (2009) 2065}
  [\href{https://arxiv.org/abs/0810.4170}{{\ttfamily 0810.4170}}].

\bibitem{1302.6994}
M.~{Takada} and W.~{Hu}, \emph{{Power spectrum super-sample covariance}},
  \href{https://doi.org/10.1103/PhysRevD.87.123504}{\emph{\prd} {\bfseries 87}
  (2013) 123504} [\href{https://arxiv.org/abs/1302.6994}{{\ttfamily
  1302.6994}}].

\bibitem{astro-ph/0307515}
G.~{Efstathiou}, \emph{{Myths and truths concerning estimation of power
  spectra: the case for a hybrid estimator}},
  \href{https://doi.org/10.1111/j.1365-2966.2004.07530.x}{\emph{\mnras}
  {\bfseries 349} (2004) 603}
  [\href{https://arxiv.org/abs/astro-ph/0307515}{{\ttfamily
  astro-ph/0307515}}].

\bibitem{1601.05779}
E.~{Krause} and T.~{Eifler}, \emph{{cosmolike - cosmological likelihood
  analyses for photometric galaxy surveys}},
  \href{https://doi.org/10.1093/mnras/stx1261}{\emph{\mnras} {\bfseries 470}
  (2017) 2100} [\href{https://arxiv.org/abs/1601.05779}{{\ttfamily
  1601.05779}}].

\bibitem{2006.00008}
A.~{Nicola}, J.~{Dunkley} and D.~N. {Spergel}, \emph{{Joint cosmology and mass
  calibration from thermal Sunyaev-Zel'dovich cluster counts and cosmic
  shear}}, \href{https://doi.org/10.1103/PhysRevD.102.083505}{\emph{\prd}
  {\bfseries 102} (2020) 083505}
  [\href{https://arxiv.org/abs/2006.00008}{{\ttfamily 2006.00008}}].

\bibitem{astro-ph/9912508}
M.~{Bartelmann} and P.~{Schneider}, \emph{{Weak gravitational lensing}},
  \href{https://doi.org/10.1016/S0370-1573(00)00082-X}{\emph{\physrep}
  {\bfseries 340} (2001) 291}
  [\href{https://arxiv.org/abs/astro-ph/9912508}{{\ttfamily
  astro-ph/9912508}}].

\bibitem{1706.09359}
E.~{Krause}, T.~F. {Eifler}, J.~{Zuntz}, O.~{Friedrich}, M.~A. {Troxel},
  S.~{Dodelson} et~al., \emph{{Dark Energy Survey Year 1 Results: Multi-Probe
  Methodology and Simulated Likelihood Analyses}}, {\emph{arXiv e-prints}
  (2017) arXiv:1706.09359} [\href{https://arxiv.org/abs/1706.09359}{{\ttfamily
  1706.09359}}].

\bibitem{1953ApJ...117..134L}
D.~N. {Limber}, \emph{{The Analysis of Counts of the Extragalactic Nebulae in
  Terms of a Fluctuating Density Field.}},
  \href{https://doi.org/10.1086/145672}{\emph{\apj} {\bfseries 117} (1953)
  134}.

\bibitem{astro-ph/0308260}
N.~{Afshordi}, Y.-S. {Loh} and M.~A. {Strauss}, \emph{{Cross-correlation of the
  cosmic microwave background with the 2MASS galaxy survey: Signatures of dark
  energy, hot gas, and point sources}},
  \href{https://doi.org/10.1103/PhysRevD.69.083524}{\emph{\prd} {\bfseries 69}
  (2004) 083524} [\href{https://arxiv.org/abs/astro-ph/0308260}{{\ttfamily
  astro-ph/0308260}}].

\bibitem{1702.05301}
M.~{Kilbinger}, C.~{Heymans}, M.~{Asgari}, S.~{Joudaki}, P.~{Schneider},
  P.~{Simon} et~al., \emph{{Precision calculations of the cosmic shear power
  spectrum projection}},
  \href{https://doi.org/10.1093/mnras/stx2082}{\emph{\mnras} {\bfseries 472}
  (2017) 2126} [\href{https://arxiv.org/abs/1702.05301}{{\ttfamily
  1702.05301}}].

\bibitem{astro-ph/0207664}
R.~E. {Smith}, J.~A. {Peacock}, A.~{Jenkins}, S.~D.~M. {White}, C.~S. {Frenk},
  F.~R. {Pearce} et~al., \emph{{Stable clustering, the halo model and
  non-linear cosmological power spectra}},
  \href{https://doi.org/10.1046/j.1365-8711.2003.06503.x}{\emph{\mnras}
  {\bfseries 341} (2003) 1311}
  [\href{https://arxiv.org/abs/astro-ph/0207664}{{\ttfamily
  astro-ph/0207664}}].

\bibitem{1208.2701}
R.~{Takahashi}, M.~{Sato}, T.~{Nishimichi}, A.~{Taruya} and M.~{Oguri},
  \emph{{Revising the Halofit Model for the Nonlinear Matter Power Spectrum}},
  \href{https://doi.org/10.1088/0004-637X/761/2/152}{\emph{\apj} {\bfseries
  761} (2012) 152} [\href{https://arxiv.org/abs/1208.2701}{{\ttfamily
  1208.2701}}].

\bibitem{0807.1733}
T.~{Matsubara}, \emph{{Nonlinear perturbation theory with halo bias and
  redshift-space distortions via the Lagrangian picture}},
  \href{https://doi.org/10.1103/PhysRevD.78.083519}{\emph{\prd} {\bfseries 78}
  (2008) 083519} [\href{https://arxiv.org/abs/0807.1733}{{\ttfamily
  0807.1733}}].

\bibitem{2012.04636}
S.-F. {Chen}, Z.~{Vlah}, E.~{Castorina} and M.~{White}, \emph{{Redshift-Space
  Distortions in Lagrangian Perturbation Theory}}, {\emph{arXiv e-prints}
  (2020) arXiv:2012.04636} [\href{https://arxiv.org/abs/2012.04636}{{\ttfamily
  2012.04636}}].

\bibitem{1712.05768}
L.~H. {Garrison}, D.~J. {Eisenstein}, D.~{Ferrer}, J.~L. {Tinker}, P.~A.
  {Pinto} and D.~H. {Weinberg}, \emph{{The Abacus Cosmos: A Suite of
  Cosmological N-body Simulations}},
  \href{https://doi.org/10.3847/1538-4365/aabfd3}{\emph{\apjs} {\bfseries 236}
  (2018) 43} [\href{https://arxiv.org/abs/1712.05768}{{\ttfamily 1712.05768}}].

\bibitem{Maksimova:2021xxx}
N.~{Maksimova}, L.~{Garrison}, D.~{Eisenstein}, B.~{Hadzhiyska} and S.~{Bose},
  \emph{{{\textsc{AbacusSummit}: A Massive Set of High-Accuracy,
  High-Resolution N-Body Simulations}}}, {\emph{In preparation} }
  [\href{https://arxiv.org/abs/2104.XXXXX}{{\ttfamily 2104.XXXXX}}].

\bibitem{Garrison:2021xxx}
L.~{Garrison}, D.~{Eisenstein}, N.~{Maksimova}, D.~{Ferrer}, B.~{Hadzhiyska},
  M.~V. {Metchnik} et~al., \emph{{{The Abacus Cosmological $N$-body Code}}},
  {\emph{In preparation} } [\href{https://arxiv.org/abs/2104.XXXXX}{{\ttfamily
  2104.XXXXX}}].

\bibitem{2019MNRAS.485.3370G}
L.~H. {Garrison}, D.~J. {Eisenstein} and P.~A. {Pinto}, \emph{{A high-fidelity
  realization of the Euclid code comparison N-body simulation with ABACUS}},
  \href{https://doi.org/10.1093/mnras/stz634}{\emph{\mnras} {\bfseries 485}
  (2019) 3370} [\href{https://arxiv.org/abs/1810.02916}{{\ttfamily
  1810.02916}}].

\bibitem{2101.11014}
N.~{Kokron}, J.~{DeRose}, S.-F. {Chen}, M.~{White} and R.~H. {Wechsler},
  \emph{{The cosmology dependence of galaxy clustering and lensing from a
  hybrid $N$-body-perturbation theory model}}, {\emph{arXiv e-prints} (2021)
  arXiv:2101.11014} [\href{https://arxiv.org/abs/2101.11014}{{\ttfamily
  2101.11014}}].

\bibitem{2005.00523}
S.-F. {Chen}, Z.~{Vlah} and M.~{White}, \emph{{Consistent modeling of velocity
  statistics and redshift-space distortions in one-loop perturbation theory}},
  \href{https://doi.org/10.1088/1475-7516/2020/07/062}{\emph{\jcap} {\bfseries
  2020} (2020) 062} [\href{https://arxiv.org/abs/2005.00523}{{\ttfamily
  2005.00523}}].

\bibitem{1807.06209}
{Planck Collaboration}, N.~{Aghanim}, Y.~{Akrami}, M.~{Ashdown}, J.~{Aumont},
  C.~{Baccigalupi} et~al., \emph{{Planck 2018 results. VI. Cosmological
  parameters}}, \href{https://doi.org/10.1051/0004-6361/201833910}{\emph{\aap}
  {\bfseries 641} (2020) A6}
  [\href{https://arxiv.org/abs/1807.06209}{{\ttfamily 1807.06209}}].

\bibitem{2021arXiv210111014K}
N.~{Kokron}, J.~{DeRose}, S.-F. {Chen}, M.~{White} and R.~H. {Wechsler},
  \emph{{The cosmology dependence of galaxy clustering and lensing from a
  hybrid $N$-body-perturbation theory model}}, {\emph{arXiv e-prints} (2021)
  arXiv:2101.11014} [\href{https://arxiv.org/abs/2101.11014}{{\ttfamily
  2101.11014}}].

\bibitem{2021arXiv210112187Z}
M.~{Zennaro}, R.~E. {Angulo}, M.~{Pellejero-Ib{\'a}{\~n}ez}, J.~{St{\"u}cker},
  S.~{Contreras} and G.~{Aric{\`o}}, \emph{{The BACCO simulation project:
  biased tracers in real space}}, {\emph{arXiv e-prints} (2021)
  arXiv:2101.12187} [\href{https://arxiv.org/abs/2101.12187}{{\ttfamily
  2101.12187}}].

\bibitem{2019PDU....24..260B}
T.~{Brinckmann} and J.~{Lesgourgues}, \emph{{MontePython 3: Boosted MCMC
  sampler and other features}},
  \href{https://doi.org/10.1016/j.dark.2018.100260}{\emph{Physics of the Dark
  Universe} {\bfseries 24} (2019) 100260}
  [\href{https://arxiv.org/abs/1804.07261}{{\ttfamily 1804.07261}}].

\bibitem{1812.05995}
N.~E. {Chisari}, D.~{Alonso}, E.~{Krause}, C.~D. {Leonard}, P.~{Bull},
  J.~{Neveu} et~al., \emph{{Core Cosmology Library: Precision Cosmological
  Predictions for LSST}},
  \href{https://doi.org/10.3847/1538-4365/ab1658}{\emph{\apjs} {\bfseries 242}
  (2019) 2} [\href{https://arxiv.org/abs/1812.05995}{{\ttfamily 1812.05995}}].

\bibitem{2011.03411}
A.~{Porredon}, M.~{Crocce}, P.~{Fosalba}, J.~{Elvin-Poole}, A.~{Carnero
  Rosell}, R.~{Cawthon} et~al., \emph{{Dark Energy Survey Year 3 results:
  Optimizing the lens sample in a combined galaxy clustering and galaxy-galaxy
  lensing analysis}},
  \href{https://doi.org/10.1103/PhysRevD.103.043503}{\emph{\prd} {\bfseries
  103} (2021) 043503} [\href{https://arxiv.org/abs/2011.03411}{{\ttfamily
  2011.03411}}].

\bibitem{astro-ph/0108151}
R.~H. {Wechsler}, J.~S. {Bullock}, J.~R. {Primack}, A.~V. {Kravtsov} and
  A.~{Dekel}, \emph{{Concentrations of Dark Halos from Their Assembly
  Histories}}, \href{https://doi.org/10.1086/338765}{\emph{\apj} {\bfseries
  568} (2002) 52} [\href{https://arxiv.org/abs/astro-ph/0108151}{{\ttfamily
  astro-ph/0108151}}].

\bibitem{astro-ph/0506510}
L.~{Gao}, V.~{Springel} and S.~D.~M. {White}, \emph{{The age dependence of halo
  clustering}},
  \href{https://doi.org/10.1111/j.1745-3933.2005.00084.x}{\emph{\mnras}
  {\bfseries 363} (2005) L66}
  [\href{https://arxiv.org/abs/astro-ph/0506510}{{\ttfamily
  astro-ph/0506510}}].

\bibitem{astro-ph/0512416}
R.~H. {Wechsler}, A.~R. {Zentner}, J.~S. {Bullock}, A.~V. {Kravtsov} and
  B.~{Allgood}, \emph{{The Dependence of Halo Clustering on Halo Formation
  History, Concentration, and Occupation}},
  \href{https://doi.org/10.1086/507120}{\emph{\apj} {\bfseries 652} (2006) 71}
  [\href{https://arxiv.org/abs/astro-ph/0512416}{{\ttfamily
  astro-ph/0512416}}].

\bibitem{1911.02610}
B.~{Hadzhiyska}, S.~{Bose}, D.~{Eisenstein}, L.~{Hernquist} and D.~N.
  {Spergel}, \emph{{Limitations to the `basic' HOD model and beyond}},
  \href{https://doi.org/10.1093/mnras/staa623}{\emph{\mnras} {\bfseries 493}
  (2020) 5506} [\href{https://arxiv.org/abs/1911.02610}{{\ttfamily
  1911.02610}}].

\bibitem{1406.4159}
D.~{Alonso}, E.~{Eardley} and J.~A. {Peacock}, \emph{{Halo abundances within
  the cosmic web}}, \href{https://doi.org/10.1093/mnras/stu2632}{\emph{\mnras}
  {\bfseries 447} (2015) 2683}
  [\href{https://arxiv.org/abs/1406.4159}{{\ttfamily 1406.4159}}].

\bibitem{1801.04878}
S.~{Alam}, Y.~{Zu}, J.~A. {Peacock} and R.~{Mandelbaum}, \emph{{Cosmic web
  dependence of galaxy clustering and quenching in SDSS}},
  \href{https://doi.org/10.1093/mnras/sty3477}{\emph{\mnras} {\bfseries 483}
  (2019) 4501} [\href{https://arxiv.org/abs/1801.04878}{{\ttfamily
  1801.04878}}].

\bibitem{oliphant2006guide}
T.~E. Oliphant, \emph{A guide to NumPy}, vol.~1. Trelgol Publishing USA, 2006.

\bibitem{van2011numpy}
S.~Van Der~Walt, S.~C. Colbert and G.~Varoquaux, \emph{The numpy array: a
  structure for efficient numerical computation}, {\emph{Computing in Science
  \& Engineering} {\bfseries 13} (2011) 22}.

\bibitem{2020SciPy-NMeth}
P.~{Virtanen}, R.~{Gommers}, T.~E. {Oliphant}, M.~{Haberland}, T.~{Reddy},
  D.~{Cournapeau} et~al., \emph{{SciPy 1.0: Fundamental Algorithms for
  Scientific Computing in Python}},
  \href{https://doi.org/https://doi.org/10.1038/s41592-019-0686-2}{\emph{Nature
  Methods} {\bfseries 17} (2020) 261}.

\bibitem{astropy:2013}
{Astropy Collaboration}, T.~P. {Robitaille}, E.~J. {Tollerud}, P.~{Greenfield},
  M.~{Droettboom}, E.~{Bray} et~al., \emph{{Astropy: A community Python package
  for astronomy}},
  \href{https://doi.org/10.1051/0004-6361/201322068}{\emph{\aap} {\bfseries
  558} (2013) A33} [\href{https://arxiv.org/abs/1307.6212}{{\ttfamily
  1307.6212}}].

\bibitem{astropy:2018}
{Astropy Collaboration}, A.~M. {Price-Whelan}, B.~M. {Sip{H{o}}cz}, H.~M.
  {G{"u}nther}, P.~L. {Lim}, S.~M. {Crawford} et~al., \emph{{The Astropy
  Project: Building an Open-science Project and Status of the v2.0 Core
  Package}}, \href{https://doi.org/10.3847/1538-3881/aabc4f}{\emph{aj}
  {\bfseries 156} (2018) 123}
  [\href{https://arxiv.org/abs/1801.02634}{{\ttfamily 1801.02634}}].

\bibitem{Zonca2019}
A.~Zonca, L.~Singer, D.~Lenz, M.~Reinecke, C.~Rosset, E.~Hivon et~al.,
  \emph{healpy: equal area pixelization and spherical harmonics transforms for
  data on the sphere in python},
  \href{https://doi.org/10.21105/joss.01298}{\emph{Journal of Open Source
  Software} {\bfseries 4} (2019) 1298}.

\bibitem{2018AJ....156..160H}
N.~{Hand}, Y.~{Feng}, F.~{Beutler}, Y.~{Li}, C.~{Modi}, U.~{Seljak} et~al.,
  \emph{{nbodykit: An Open-source, Massively Parallel Toolkit for Large-scale
  Structure}}, \href{https://doi.org/10.3847/1538-3881/aadae0}{\emph{\aj}
  {\bfseries 156} (2018) 160}
  [\href{https://arxiv.org/abs/1712.05834}{{\ttfamily 1712.05834}}].

\bibitem{2019arXiv191013970L}
A.~{Lewis}, \emph{{GetDist: a Python package for analysing Monte Carlo
  samples}}, {\emph{arXiv e-prints} (2019) arXiv:1910.13970}
  [\href{https://arxiv.org/abs/1910.13970}{{\ttfamily 1910.13970}}].

\bibitem{Hunter:2007}
J.~D. Hunter, \emph{Matplotlib: A 2d graphics environment},
  \href{https://doi.org/10.1109/MCSE.2007.55}{\emph{Computing in Science \&
  Engineering} {\bfseries 9} (2007) 90}.

\end{thebibliography}\endgroup

\end{document}